\documentclass[preprint, longbibliography, 10pt, twocolumn, floatfix]{revtex4-1} 
\usepackage{dcolumn}
\usepackage{textcomp,gensymb} 
\usepackage{amssymb}
\usepackage{amsmath}
\usepackage{mathrsfs}
\usepackage{bm}
\usepackage[dvipsnames]{xcolor}

\usepackage{relsize}

\usepackage{media9} 
\usepackage[hidelinks]{hyperref}

\usepackage{graphicx}
\usepackage{color}
\usepackage{xcolor}
\definecolor{red}{rgb}{1,0,0}
\definecolor{blue}{rgb}{0,0,1}

\graphicspath{{img/}}
\renewcommand{\figurename}{\textbf{Figure}}

\newcommand{\uvec}[1]{\hat{\mathbf{#1}}}

\newcommand{\sref}[1]{S\ref{#1}}

\usepackage{multirow}
\usepackage{siunitx}
	\usepackage{float}

\usepackage{xr}

\makeatletter
\newcommand*{\addFileDependency}[1]{
  \typeout{(#1)}
  \@addtofilelist{#1}
  \IfFileExists{#1}{}{\typeout{No file #1.}}
}
\makeatother

\begin{document}

\title{Twist dynamics and buckling instability of ring DNA: Effect of groove asymmetry and anisotropic bending}

\author{Yair Augusto Guti\'{e}rrez Fosado$^{1,*}$, Fabio Landuzzi$^{1,*}$ and Takahiro Sakaue$^{1,2,\dagger}$}
\affiliation{$^1$ Department of Physics and Mathematics, Aoyama Gakuin University
5-10-1 Fuchinobe, Chuo-ku, Sagamihara-shi, Kanagawa 252-5258, JAPAN. $^2$ PRESTO, Japan Science and Technology Agency (JST), 4-1-8 Honcho Kawaguchi, Saitama 332-0012, Japan. $^*$ Joint first author \\ $^\dagger$For correspondence: T. Sakaue (sakaue@phys.aoyama.ac.jp)}


\begin{abstract}
\textbf{By combining analytical theory and Molecular Dynamics simulations we study the relaxation dynamics of DNA circular plasmids that initially undergo a local twist perturbation. We identify three distinctive time scales; ($\mathcal{I}$) a rapid relaxation of local bending, ($\mathcal{II}$) the slow twist spreading, and ($\mathcal{III}$) the buckling transition taking place in a much longer time scale. In all of these stages, the twist-bend coupling arising from the groove asymmetry in DNA double helix clearly manifests. In particular, the separation of time scales allows to deduce an effective diffusion equation in stage ($\mathcal{II}$), with a diffusion coefficient influenced by the twist-bend coupling. We also discuss the mapping of the realistic DNA model to the simplest isotropic twistable worm-like chain using the renormalized bending and twist moduli; although useful in many cases, it fails to make a quantitative prediction on the instability mode of buckling transition.
}
\end{abstract}

\maketitle


\emph{\label{sec:introduction}Introduction --} It has become increasingly evident that not only the information encoded in the DNA sequence is relevant in several biological processes, but also that the elastic properties of DNA and its topology play a key role in its functioning~\cite{BrackleyE3605,MolBioCell,2009Cook}. In the transcription process, for example, the RNA polymerase locally reshapes DNA as it reads the sequence along it. This local deformation generates stress of the helix that dynamically drives overtwisting ahead and undertwisting behind the polymerase~\cite{Liu1987TSD}. It has been hypothesized that this stress could in principle influence the dynamics of nucleosomes, the binding of proteins along the DNA, the gene expression, among other regulatory processes~\cite{BARANELLO2012,Brackley2016,Kouzine2013}. Thus changing the role we perceive DNA from a passive entity that storage information to an active participant of the regulation of gene activity.

To address the elastic response of DNA to mechanical manipulations, the isotropic twistable worm-like-chain (TWLC) model is usually employed. This model describes a double-stranded (ds) DNA helix as an inextensible and isotropic elastic rod characterized by only two elastic constants: the bending stiffness ($A$) and the torsional stiffness ($C$). However, the actual DNA is equipped with hard and soft directions for bending (anisotropy), and furthermore, the geometrical asymmetry imposed by the presence of the minor and major grooves of the dsDNA helix gives rise to a coupling between twist and bend~\cite{1994MarkoSiggia}. Recent studies have revealed that these elements, not included in the isotropic TWLC, are relevant to the DNA physics in several contexts, including the bending and twisting of DNA in the nucleosome scale~\cite{2018Enricotwistwavesnuc,2019Ecarlonpolygonalshapes,2008GolestanianPRL}.

\begin{figure*}[htpb]
\centering
\includegraphics[width=1.0\textwidth]{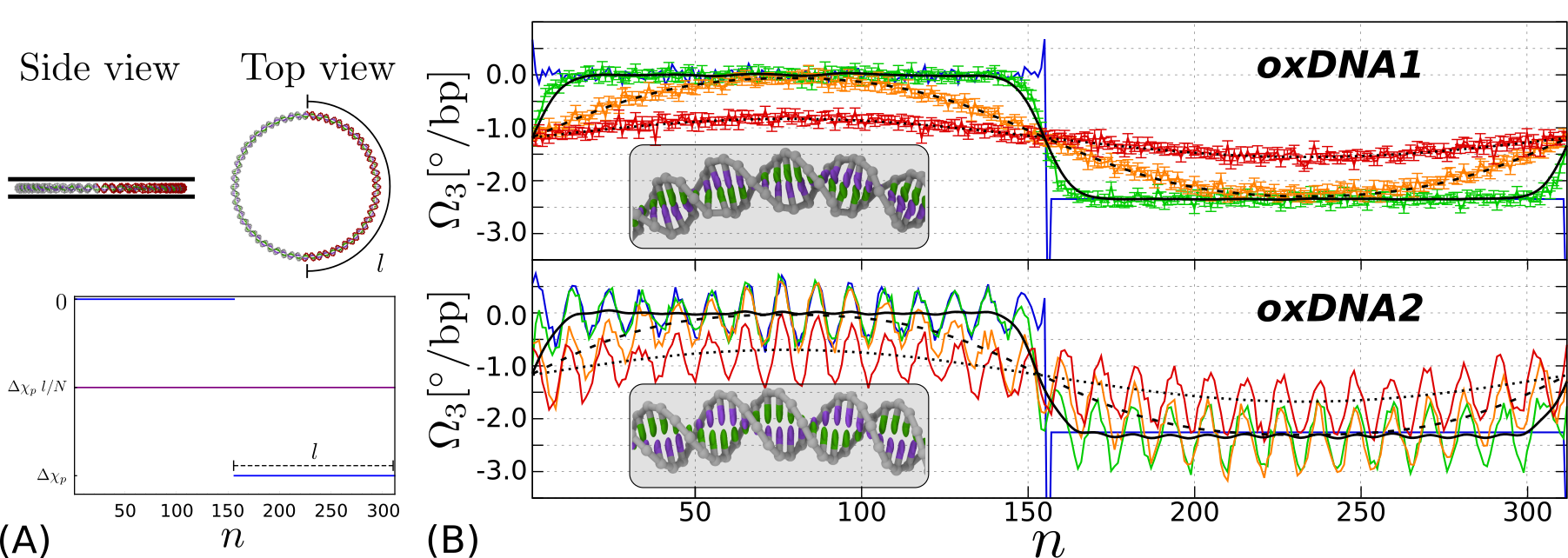}
\caption{Twist diffusion of undertwisted rings. (\textbf{A}) Schematic of twist perturbation protocol. Initially, a segment of the DNA ($l$ out of total $N$ base pairs ) is constrained to have a twist density $\Delta \chi_p$ (blue line in the plot). After the constraint is released at $t=0$, the system relaxes to a state of uniform excess twist density $\Omega_3 = \Delta \chi_p l/N$ (purple line in the plot). (\textbf{B}) Local twist deformation $\Omega_{3}$ as function of the contour length for oxDNA1 (top) and oxDNA2 (bottom). Color lines (blue, green, yellow and red) represent results at different times ($t=0, 5, 2500$ and $15000 \tau_{\text{LJ}}$). Black lines (solid, dashed and dotted) are solutions to the diffusion equation at the corresponding times for $t>0$ (see \cite{SM} for details). The insets show sections in the relaxed part of the ring before the release of under-twist. The overtwisted case is also presented in~\cite{SM}.
}
\label{fig:paneltwdiffusion}
\vspace{-0.5cm}
\end{figure*}

Despite its importance, the study in this direction has so far been restricted to the statics, which thus does not resolve important time-dependent processes, such as how efficiently torsional stress is transported to remote DNA. Here we try to shed some light onto this subject by combining an analytical theory and Molecular Dynamic (MD) simulations. The protocol we employ in this study is shown in Fig.~\ref{fig:paneltwdiffusion}
, which allows us to investigate the rate at which the stress can be relief through the propagation of an over/under twisted region of a short DNA ring to the adjacent base-pairs.

\emph{\label{sec.background} Reminder on the model --}
Let us describe the center line of the DNA as a curve, $\mathbf{r}(s)$, parameterized by arc length ($s$) and with a fixed total length $L$. Its conformation is completely characterized by the set of orthonormal vectors $\{\uvec{e}_{1}(s), \uvec{e}_{2}(s), \uvec{e}_{3}(s)\}$ that define a local reference frame at each point $s$ on the curve. The tangent vector to the center line is $\uvec{e}_{3}=\frac{d}{ds} \mathbf{r}(s)$ and by convention, $\uvec{e}_{1}$ lies in the symmetry axis of the grooves in the direction of the major groove and $\uvec{e}_{2}= \uvec{e}_{3}  \times  \uvec{e}_{1}$.
Associated with the conformation, there is a local strain $\mathbf{\Omega}(s) = \Omega_{1}(s) \uvec{e}_{1}(s) + \Omega_{2}(s) \uvec{e}_{2}(s) + \Omega_{3}(s) \uvec{e}_{3}(s) $, that connects two reference frames located at $s$ and $s+ds$. This satisfies the following differential equation.

\begin{equation}
\frac{d \uvec{e}_{i}}{ds} = [\mathbf{\Omega} + \chi_{0} \uvec{e}_{3}(s)]  \times \uvec{e}_{i} (s),
\label{eq.deds}
\end{equation}

\noindent with $i=1,2,3$ and $\chi_{0} ( \simeq 1.75 \ {\rm nm}^{-1})$ the intrinsic twist density of the DNA helix. The three components of the deformation vector are $\Omega_{i}(s) = \mathbf{\Omega}(s) \cdot \uvec{e}_{i}(s)$, where $\Omega_{3}$ is the local excess (or deficit) of twist density with respect to $\chi_{0}$. On the other hand, $\Omega_{1}$ and $\Omega_{2}$ represent the bending densities related to the tilt and roll degrees of freedom of the dsDNA.

The elastic free energy functional in terms of the local deformations truncated at quadratic
order can be written~\cite{1994MarkoSiggia} as $E  = \int_{0}^{L} \varepsilon_0 (\mathbf{\Omega}) ds $ with
\begin{eqnarray}
\varepsilon_0 (\mathbf{\Omega}) = \frac{1}{2}( A_{1} \Omega_{1}^{2} + A_{2} \Omega_{2}^{2} + C \Omega_{3}^{2} + 2G \Omega_{2} \Omega_{3}),
\label{eq.energydensitytwlc}
\end{eqnarray}

\noindent where the bending rigidities about the axes $\uvec{e}_{1}$ and $\uvec{e}_{2}$ are $A_{1}$ and $A_{2}$, respectively. $C$ expresses the torsional stiffness and $G$ represents the twist-bend coupling between the local deformations $\Omega_{2}$ and $\Omega_{3}$. These parameters, in general, depend on the sequence of the molecule~\cite{2017EnricoEoGA} and therefore on $s$, but for simplicity here we only consider homopolymers for which they are constant.

Important features of the model become transparent if the energy density is transformed into a complete square as
\begin{eqnarray}
&&\varepsilon_0 (\mathbf{\Omega}) =\frac{A_{1}}{2}  \Omega_{1}^{2} +\frac{A_{2}}{2} \left( \Omega_{2} + \frac{G}{A_2} \Omega_3 \right)^{2} + \frac{{\tilde C}}{2} \Omega_{3}^{2},
\label{e_complete_square}
\end{eqnarray}
\noindent where ${\tilde C} = C \left( 1- \frac{G^2}{A_2 C}\right)$ is the renormalized twist modulus. The consequence of this renormalization is evident after integrating out the bending degrees of freedom from the weight $P(\{ \Omega_i \}) \propto e^{-\beta \varepsilon_0(\{\Omega_i \})}$~\cite{SM}, implying the softening of twisting response. Similarly, the bending about the $\uvec{e}_{2}$ axis modulus is affected and it can be conveniently written as ${\tilde A}_2 = A_2 \left( 1- \frac{G^2}{A_2 C}\right)$.

The bending ($l_{b}$) and torsional ($l_{\tau}$) persistence lengths that are commonly defined by the correlation of the reference frames along $s$, can be expressed in terms of these renormalized elastic constants~\cite{Nomidis2017}: ${l_{b}} = \beta {\tilde A}$ and $l_{\tau} = 2\beta\tilde{C}$ where ${\tilde A}=\frac{2}{A_1^{-1} + {\tilde A}_2^{-1}}$ is the harmonic mean of $A_1$ and ${\tilde A}_2$.

\emph{\label{sec.closed} Closed configurations --} The elastic free energy functional in terms of the local deformations has been thoroughly studied in the past for linear~\cite{1994MarkoSiggia,2017EnricoEoGA} and ring~\cite{2018Enricotwistwavesnuc,2019Ecarlonpolygonalshapes} molecules. Following ref.~\cite{2019Ecarlonpolygonalshapes}, here we employ the energy density $\varepsilon$ for a torsionally constraint DNA ring

\begin{equation}
\begin{aligned}
\varepsilon(\mathbf{\Omega}) = \varepsilon_0 (\mathbf{\Omega}) - \mu [\Omega_{1} \sin (\chi s) + \Omega_{2} \cos (\chi s)] - \lambda \Omega_3,
\end{aligned}
\label{eq.energydensity}
\end{equation}
\noindent where the last two terms with Lagrange multipliers $\mu$ and $\lambda$ are introduced to represent the ring closure constraint, which allows to analytically describe the minimum energy configuration of a ring~\cite{2019Ecarlonpolygonalshapes}.
The first term enforces the bending ($\mathbf{\Omega}_{b}=\mathbf{\Omega}_{1} + \mathbf{\Omega}_{2}$) to take place along the unitary vector $\uvec{x}=\sin (\chi s) \uvec{e}_{1} + \cos (\chi s) \uvec{e}_{2}$ pointing in the direction perpendicular to the plane spanned by the molecule, while the second term accounts for the presence of twist excess $\Delta \chi = \chi - \chi_0$. 

Notably, the renormalized elastic constants enter in the equations that identify the ground state of ring DNA molecules. This is found by the minimization of the energy density with respect to $\Omega_{i}$. Thus,  the elastic strains at ground state for a torsionally stressed planar ring are

\begin{eqnarray}
\Omega_{1}^{0}(s)&=&\frac{\mu}{A_{1}} \sin(\chi s), \nonumber \\ 
\Omega_{2}^{0}(s)&=&\frac{\mu}{\tilde{A}_{2}} \cos(\chi s) - \frac{\lambda G}{{\tilde C}  A_2}, \label{eq.groundstate} \\ \Omega_{3}^{0}(s)&=&-\frac{\mu G}{C {\tilde A}_2}\cos{(\chi s)}+ \frac{\lambda}{{\tilde C}}. \nonumber
\end{eqnarray}

\noindent where $\mu=k_{B}T l_{b}/R_{0}$ is identified with the external bending torque necessary for a ring with a bending persistence length $l_{b}$ to adopt a configuration with radius  $R_{0}$~\cite{2018Enricotwistwavesnuc}, while $\lambda = {\tilde C} \Delta \chi$ is the external twisting torques as inspected from the relation $\int_0^L ds \ \Omega_3^0(s) \ (= \Delta \chi L) =  \lambda L/{\tilde C}$.
The oscillation of $\Omega_1^0$ and $\Omega_2^0$ is a natural consequence of the DNA helical structure. Marked features here are: (i) the anisotropy ($A_1 \neq {\tilde A}_2$) implies unequal bending amplitudes, leading to the non-constant curvature $\kappa(s)= \sqrt{(\Omega_1^0(s))^2 + (\Omega_2^0(s))^2}$; (ii) the twist-bend coupling ($G > 0$) induces a ``twist wave", i.e., a periodic modulation in $\Omega_3^0(s)$ which is in anti-phase with $\Omega_2^0(s)$; and (iii) it produces a constant shift in $\Omega_2^0(s)$ and $\Omega_3^0(s)$. It is also worth mentioning that the ground state energy
\begin{eqnarray}
\oint ds \ \varepsilon (\Omega_{1}^{0}(s), \Omega_{2}^{0}(s), \Omega_{3}^{0}(s)) = \left( \frac{1}{2}\frac{{\tilde A}}{R_0^2} + \frac{1}{2} {\tilde C}(\delta \chi)^2 \right),
\label{Energy_WLC_map}
\end{eqnarray}
is formally identical to that of the isotropic TWLC ring (with constant radius of curvature $R_0$) with bending and twist moduli ${\tilde A}$ and ${\tilde C}$, respectively. Although at first sight it is not obvious how the nontrivial structural properties discuss here affect the twist dynamics in real DNA, in the following we attempt to get some insight onto this.

\emph{\label{sec:results}Coarse-grained simulations of DNA --} Here we study the dynamics of twist in the oxDNA~\cite{oxDNA2010,Ouldridge:2011}, a coarse-grained model that is based on force fields tuned to  account for several geometrical and thermodynamic features of single and double stranded DNA (in its B form). One important feature of the oxDNA model for the current study is that two parameterizations are available, namely, oxDNA1 and oxDNA2. While the former describes dsDNA as a molecule with symmetric grooves, the latter introduces the appropriate groove asymmetry found in real DNA. Therefore, we expect that there is a direct mapping between these models and the theory described in the previous section. The oxDNA1 resembles the anisotropic TWLC model ($G=0, C>0 \text{ and } A_{1} \neq A_{2}$) and the oxDNA2 the more general case in which all the stiffness parameters play a role in the description.

We first investigate the diffusion of twist by performing coarse-grained molecular dynamics simulations of dsDNA mini-circles with a total length of $N=312$ bp. We follow a similar protocol of that on reference \cite{fosado2019transcriptiondriven}. The molecule is initialized with the mean distance between successive base-pairs $a=0.34$ nm. The local twist in half of the ring is set to its natural value: $a\chi_{0} = 34.5\degree$ and $34.1\degree$ for oxDNA1 and oxDNA2, hence $\Omega_3(s, t=0)=0$ for $s \in[1, l]$, with $l=N/2$. In the other half, a deficit (or excess) of one helical turn is introduced such that $\Delta \chi_p \equiv \Omega_3(s, t=0)=- 2\pi/la$ for $s \in(l, N]$. During equilibration, the undertwisted segment of the ring is constraint so the local twist is fixed and the simulation is run for $10^{5} \; \tau_{\text{LJ}}$ (simulation time) at a low temperature of 15 K. After this stage, the constraint is released at $t=0$ and we study the twist relaxation by monitoring its local value along the molecule. During the whole simulation, the system is confined in between two parallel planes to prevent writhe formation and in this way being able to study pure twist dynamics. We will show later that essentially the same result is obtained even without the confining walls. The exact same protocol was applied for both, the oxDNA1 and oxDNA2 models (see \cite{SM} for details).

We found that the twist evolution can be fitted by the solution of the diffusion equation with the appropriate initial and boundary conditions (see Fig.~\ref{fig:paneltwdiffusion}). From which we extract the diffusion coefficients $D_{I}^{-} = 0.222\pm0.009$ and $D_{II}^{-}=0.183\pm0.007 \; bp^{2}/\tau_{\text{LJ}}$ for the oxDNA1 and oxDNA2 models, respectively. Essentially the same values, $D_{I}^{+} = 0.231\pm0.002$ and $D_{II}^{+}=0.183\pm0.002 \; bp^{2}/\tau_{\text{LJ}}$, were obtained for overtwisted DNA.

\emph{\label{sec.dynamics} Dynamical equation --} 
To discuss the dynamics, we assume that the local reference frame, $\uvec{e}_{i} (s, t)$ and the strain $\Omega_i(s, t)$ are functions of both position $s$ and time $t$. The dynamical equation can be derived from the compatibility relation between the strain and the angular velocity together with the force and torque balance equations~\cite{2010Powersdynoffilaments}.
By focusing on the twist component, the compatibility relation leads to
\begin{equation}
\frac{\partial \Omega_{3}(s, t)}{\partial t} = \frac{\partial \omega_3(s, t)}{\partial s} + \left( \uvec{e}_{3} \times \frac{\partial \uvec{e}_{3}}{\partial s} \right) \cdot \frac{\partial \uvec{e}_{3}}{\partial t},
\label{eq.twconservation}
\end{equation} 
\noindent where $\omega_3(s, t)$ is the rotational rate of the curve at point $s$ at time $t$.
The torque balance along $\uvec{e}_{3}$ is
\begin{equation}
\begin{aligned}
\gamma_{r} \omega_3 =&{} \Omega_{1} M_2 -  \Omega_{2} M_{1}  + \frac{\partial M_3}{\partial s},
\label{eq.torquebalancetan}
\end{aligned}
\end{equation}
\noindent with the rotational friction coefficient $\gamma_{r}$ and the three components $M_{i} = \frac{\delta \varepsilon}{\delta \Omega_{i}}$ of the internal torque $\mathbf{M} = \sum_{i=1}^3 M_{i} \uvec{e}_{i}$, which are written in terms of the deviations $\delta \Omega_{i} = \Omega_{i} - \Omega_{i}^{0}$~\cite{SM}:
\begin{equation}
\begin{aligned}
M_{1} &={} A_{1}\delta \Omega_{1}, \\
M_{2} &={} A_{2}\delta \Omega_{2} + G\delta \Omega_{3},\\
M_{3} &={} C    \delta \Omega_{3} + G\delta \Omega_{2}.
\end{aligned}
\label{eq.torque}
\end{equation}

For an isotropic open TWLC ($A_1 = A_2$, $G=0$, $\Omega_i^{0}=0$), nonlinear terms in Eq.~(\ref{eq.torquebalancetan}) cancel out, reducing to a linear constitutive relation $\gamma_{r} \omega_3 = C \partial_s (\delta \Omega_3)$. Hence, assuming in-plane motion and deformation, Eq.~(\ref{eq.twconservation}) indicates the diffusive transport of the excess twist density $\Omega_3(s, t)$ with the diffusion coefficient $C/\gamma_r$. 

In our more general model, however, the story looks more complicated. A key observation here is that there is a conserved quantity due to the topological constraint $\mathrm{Lk} = \mathrm{Tw} + \mathrm{Wr}$. For a planar ring, the writhing number $\mathrm{Wr}$ is zero and the invariance of the linking number $\mathrm{Lk}$ implies that the total twist $\mathrm{Tw} = \oint ds \ [\chi_0 + \Omega_3(s,t)] $ is conserved. Therefore, there is a slow variable $\delta \Omega_3(s,t)$ associated to the twist relaxation process. In fact, Eq.~(\ref{eq.twconservation}) represents the conservation law of $\Omega_{3}$, where the rotation rate $\omega_3$ is regarded as a twist current and the last term acts as a source of twist that comes from the out of plane deformations~\cite{2010Powersdynoffilaments,1998Kamien,2011Wada}.
Thus, at each moment, the local bending strains $\delta \Omega_1(s,t), \delta \Omega_2(s,t)$ are quickly equilibrated to the state given by $M_1=M_2=0$, with which the twist strain $\delta \Omega_3(s,t)$ evolves over a longer time scale. Note that the above conditions on $M_1$, $M_2$ are equivalent to finding the averages $\langle \delta \Omega_1 \rangle$, $\langle \delta\Omega_2 + G\delta\Omega_3/A_2 \rangle$ through the integration of the bending degrees of freedom~\cite{SM}, which indicates that the renormalized modulus ${\tilde C}$ plays a role in twist dynamics. We confirm this by plugging Eq.~(\ref{eq.torquebalancetan}) into Eq.~(\ref{eq.torque}) with $M_1=M_2=0$ and finding $\gamma_{r} \omega_3 = {\tilde C} \partial_s (\delta \Omega_3)$. This result, combined with Eq.~(\ref{eq.twconservation}) leads to the diffusion equation of twist
\begin{equation}
\frac{\partial \delta \Omega_{3}}{\partial t} = \tilde{D} \frac{\partial^{2} \delta \Omega_{3} }{\partial s^{2}},
\label{eq.twdiffusion}
\end{equation}
\noindent with the diffusion coefficient $\tilde{D} = \tilde{C}/ \gamma_{r}$ and where we neglect the contributions of the twist source term related to the out of plane deformations. 

Thereby we predict that for two similar DNA molecules, one with symmetric grooves and the other with the usual asymmetry, the latter (with lower $\tilde C$) will exhibit a slower twist diffusion. 
From the elastic parameters obtained in reference~\cite{2017EnricoEoGA} and reported in \cite{SM}, we compute the rescaled twist modulus $\beta\tilde{C}_{I}=77$ nm and $\beta\tilde{C}_{II}\simeq 61$ nm for oxDNA1 and oxDNA2, respectively. Remarkably, the ratio between these two quantities ($\tilde{C}_{II}/\tilde{C}_{I}=0.79$) is in excellent agreement with the ratio of the twist diffusion coefficients found in our simulations ($D_{II}/D_{I}=0.81\pm0.02$), obtained by averaging the results from over and undertwisted rings. 

Moreover, our argument ($M_{2}=0$) suggests that the twist diffusion in DNA is followed by a {\it bend diffusion} that is induced by the twist-bend coupling. Figure~\sref{fig:paneldefevolution}(E) demonstrates that this is indeed the case for oxDNA2, where $\Omega_2$ relaxes diffusively following the behavior of $\Omega_3$ with basically the same diffusion coefficient. Such a phenomenon is not expected and hence not observed in the oxDNA1 without groove asymmetry.

Finally, we note that since the diffusion coefficient in Eq.~(\ref{eq.twdiffusion}) only depends on the elastic parameters and does not depend explicitly on the temperature of the system, we expect a similar behavior at room temperature (the temperature, though, affects the persistence lengths $l_{\tau}$ and $l_{b}$). We corroborate this behavior by running simulations with the same protocol described above at $T=300K$~\cite{SM}.

\begin{figure}[htp]
\centering
\includegraphics[width=0.45\textwidth]{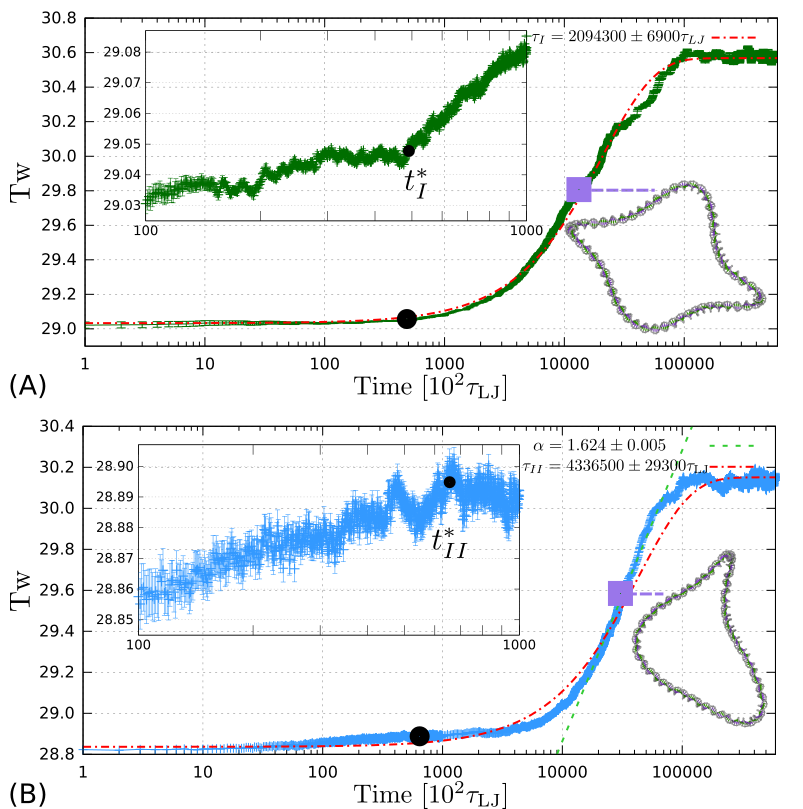}
\caption{Time evolution of the total twist ($\mathrm{Tw}$) for undertwisted oxDNA1 (\textbf{A}) and oxDNA2 (\textbf{B}). The time axis is displayed in a logarithm scale. Red dashed lines represent an exponential fit to the data $\mathrm{Tw}(t) = c1-c2*\text{exp}(-t/\tau)$. The value of the time-constant $\tau$ obtained from the fit to the oxDNA1 and oxDNA2 model are $2.09\times 10^{6}\tau_{\text{LJ}}$, $4.33\times 10 ^{6}\tau_{\text{LJ}}$. Purple squares show the timestep at which the buckling of the system (shown in the snapshots) take places. Insets show the zoom-in at short timescales. The time, $t^{*}$, of twist diffusion across the entire ring is depicted with a black dot.}
\label{fig:paneltwvstunder}
\vspace{-0.4cm}
\end{figure}

\emph{\label{sec:buckling}Buckling instability --}
We investigate the dynamics of the same system without confining planes. Figure~\ref{fig:paneltwvstunder} shows the time evolution of the total twist ($\mathrm{Tw}$), where we observe that at short times (smaller than $10^{5} \; \tau_{\text{LJ}}$) the change in $\mathrm{Tw}$ is insignificant and in consequence that the out of plane deformations of the ring are negligible. Furthermore, we expect that the time required for the deficit of twist to diffuse across the entire ring, a distance of $N-l \  (= N/2)$ bp, is given by the relation $t^{*} \approx (N-l)^{2}/ 2\tilde{D}$. Since the value of $t^{*}$ is $t_{I}^{*}=5.53\times10^{4}$ and $t_{II}^{*}=6.76\times10^{4} \; \tau_{\text{LJ}}$ for the oxDNA1 and oxDNA2 models, respectively, our assumption of neglecting the last term in Eq.~(\ref{eq.twconservation}) during the diffusion stage seems to be appropriate even after getting rid of the confining planes.

The contour length dependence of the local deformations at a fixed time from the data in Figure~\ref{fig:paneltwvstunder} is shown in the supplementary Movies \href{run:./video/ox2_N312_deltalk1.mp4}{S1} and \href{run:./video/ox1_N312_deltalk1.mp4}{S2}. The results at the end of the diffusion process ($t \sim t_{II}^{*}$ and $t \sim t_{I}^{*}$, respectively) are comparable to those on \cite{2019Ecarlonpolygonalshapes}. This is, the local deformations display the features (i-iii) predicted in Eq.~(\ref{eq.groundstate}). As the simulation continues and $t>t^{*}$, the magnitude of the shift in $\Omega_{3}$ (and also in $\Omega_{2}$ for oxDNA2 due to nonzero $G$, see Eq.~(\ref{eq.groundstate})) decreases approaching to zero and the interchange between twist and writhe takes place. This is indicated by the rapid increase of $\mathrm{Tw}$ shown in Fig.~\ref{fig:paneltwvstunder}. The buckling of the molecule is reflected in the large scale ($\gg 2\pi / \chi_{0}$) bending deformations (see supplementary movies). The number of local minima (maxima) found in the envelope of the bending deformations, referred as the bending mode number $m$, quantifies the number of times that the ring bends back and forth across its contour length. In Fig.~\sref{app:mvst} we show the value of $m$ computed from simulations as a function of time.

From the discussion so far, one may expect that the behavior of the DNA model can be mapped to that of an isotropic TWLC using the renormalized moduli ${\tilde A}$ and ${\tilde C}$. We now show that when describing the buckling of DNA, this naive expectation only holds in a qualitative level, but fails to explain the quantitative aspects. The linear stability analysis for the isotropic TWLC ring predicts that the most unstable mode ($m^{*}$) depends on the ratio of the twist and bend elastic moduli $C/A$ and the excess linking number $\Delta \mathrm{Lk}$ such that the smaller the ratio $C/A$, the smaller the selected mode number $m^*$ at a fixed $\Delta \mathrm{Lk}$ (see Fig.~\sref{fig:panelbendingmodes})~\cite{1985Tanaka}. This mode will grow faster than the others, and therefore will be the first observed at the onset of the buckling.

\begin{figure}[thp] 
\centering
\includegraphics[width=0.475\textwidth]{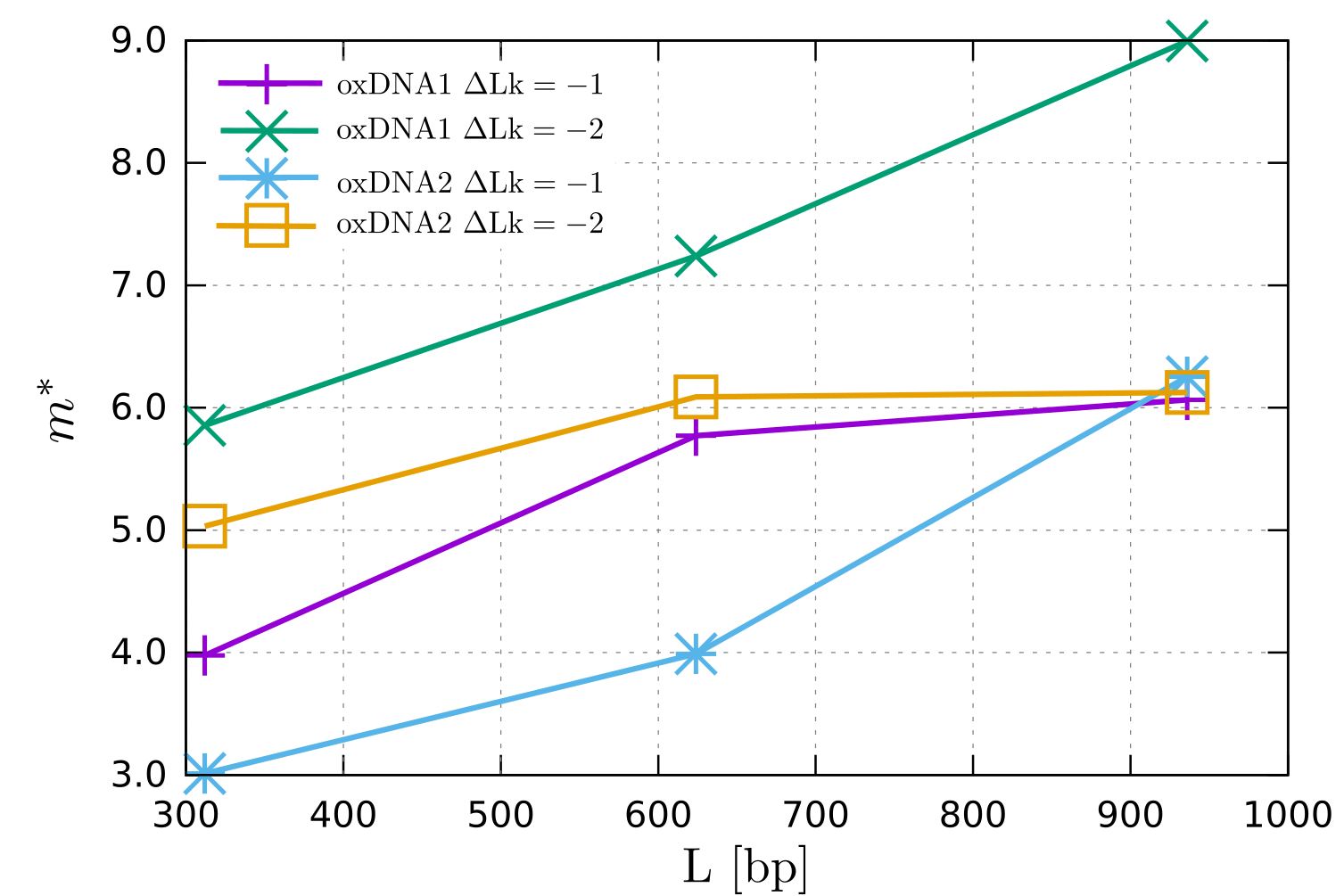}
\caption{Bending mode ($m^{*}$) selected by the system as function of the total length of the ring.}
\label{app:mvsl}
\vspace{-0.4cm}
\end{figure}

We show in Fig.~\ref{app:mvsl} the most unstable mode $m^*$ obtained from simulations as function of the DNA size. For an $N=312$ bp ring with $\Delta \mathrm{Lk} =-1$, we find $m^* \simeq 4$ for oxDNA1 and $m^* \simeq 3$ for oxDNA2 (see snapshots in Fig.~\ref{fig:paneltwvstunder}). Although these results are in good agreement with the mapping mentioned above (see \cite{SM}), this should not be regarded as a complete success. Remarkably, we numerically find that $m^*$ depends on the ring length, a feature absent in the linear theory of isotropic TWLC. On the other hand, our results also suggest a satisfactory agreement in a more qualitative level; the ratio ${\tilde C}/{\tilde A}$ of oxDNA2 $(1.6)$ is smaller than the one for oxDNA1 $(2.1)$, and as expected, so is the observed $m^*$.

The main reason for the discrepancy may lie in the helical nature of DNA (intrinsic twist $\chi_0$) and anisotropic bending. For a ring made from isotropic and untwisted filament, $A_{1}=A_{2}$ and $\chi_0=0$, the configuration that minimizes the bending energy is the one of a planar circle with constant curvature. However, non-zero $\chi_0$ leads to the periodic variation of bending strain $\Omega_1(s)$ and $\Omega_2(s) $ along the contour.  In DNA, coupled with this is the bend anisotropy, which results in preferential bending along the soft axis. As a result, a slightly off-planar configurations with varying curvature are explored in order to minimize the bending energy. These inhomogeneity may likely provide a ``hot spot" for the buckling instability. The possibility of incommensurate periodicity between the unstable mode and the underlying bending oscillation may further complicate the situation. The amplitude of the bending oscillation tends to decrease with ring size (Eq.~\ref{eq.groundstate}), but at the same time, the effect of thermal fluctuation becomes more relevant, which tends to smooth the transition.

\emph{\label{sec:conclusion}Closing Remarks --}
Through a careful numerical and theoretical analysis, we have identified three distinctive time scales in the relaxation dynamics after a local twist perturbation introduced in a torsionally constrained DNA ring. ($\mathcal{I}$) In the fastest scale, the local bending relaxation takes place given the twist strain at that moment. ($\mathcal{II}$) The twist diffusive relaxation proceeds more slowly governed by the conservation law with a topological origin. ($\mathcal{III}$) After the completion of twist diffusion, the remanent twist stress, if sufficiently strong, makes the whole ring undergo a buckling transition in a much longer time scale. It is this time scale separation that enables us to reduce the original nonlinear time evolution equation to the simple diffusion equation in the twist relaxation process ($\mathcal{II}$). It also provides rich physical insights on how the molecular features of DNA, and in particular the groove asymmetry, plays a role in the twist dynamics. Here, we have shown that the twist diffusion coefficient is entirely determined by the renormalized twist modulus ${\tilde C}$, which is smaller than the bare modulus $C$ due to the groove asymmetry induced by the twist-bend coupling. We have also predicted and verified that one component of bend ($\Omega_2$) exhibits the concomitant diffusive relaxation. The occurrence of this bend diffusion, which is tightly coupled with the twist diffusion, is again due to the groove asymmetry.

Although recent works on statics~\cite{Nomidis2017} suggest that the more realistic model reflecting molecular details of the DNA double-helix can be mapped to a simple isotropic TWLC by using the renormalized bending and twisting moduli ${\tilde A}$, ${\tilde C}$, and our present results suggest its applicability also to dynamics, we have shown that such a mapping is not almighty. A concrete counter-example is the twist-buckling, for which the instability mode can be predicted only qualitatively, but not quantitatively.

\begin{acknowledgements}
This work was supported by JSPS KAKENHI (No. JP18H05529) from MEXT, Japan, and JST, PRESTO (JPMJPR16N5). 
\end{acknowledgements}
\vspace*{-0.5 cm}

\bibliographystyle{apsrev4-1}
\bibliography{dna_twdiffbib}

\setcounter{equation}{0}
\renewcommand{\theequation}{S\arabic{equation}}
\setcounter{section}{0}
\setcounter{figure}{0}

\renewcommand{\thesection}{\Roman{section}}
\renewcommand{\figurename}{Fig.~S}
\renewcommand{\tablename}{Table~S}

{\large \bf Supplementary Material}

\setcounter{secnumdepth}{1}

\section{\label{App.} Statistical mechanics of the TWLC}
Here we use standard methods of statistical mechanics to get some important relations for the analysis presented in the main text. To this end, we first write the discretized free energy functional ($E  = \int_{0}^{L} \varepsilon_0 (\mathbf{\Omega}) ds $) at the base-pair level, with $\varepsilon_0$ given by Eq.~(\ref{eq.energydensitytwlc}): 

\begin{equation}
\begin{aligned}
E={}& \frac{a}{2}\sum_{n=1}^{N} [A_{1}\Omega_{1}^{2}(n) + A_{2}\Omega_{2}^{2}(n) + C \Omega_{3}^{2}(n) \\
& +2GC\Omega_{2}(n)\Omega_{3}(n)] \\
 ={}& \frac{a}{2}\sum_{n=1}^{N} [A_{1}\Omega_{1}^{2}(n) + A_{2}[\Omega_{2}(n) + \frac{G}{A_2} \Omega_3(n)]^{2}\\
&+ {\tilde C} \Omega_{3}^{2}(n)],
\end{aligned}
\end{equation}

\noindent where $\Omega_{i}(n)$ with $i=1,2,3$, represents the local deformation $i$ at position $n$. The second equality is obtained after using the complete square free energy density of Eq.~(\ref{e_complete_square}). Therefore, the probability that the system is in a state with energy $E$ is given by:

\begin{equation}
P = \frac{e^{-\beta E}}{Z},
\label{eq.app.prob}
\end{equation} 

\noindent where $Z$ is the partition function:

\begin{equation}
Z=\mathlarger{\int} \mathscr{D}[\mathbf{\Omega}] e^{-\beta E},
\label{eq.app.partfunc}
\end{equation}

\noindent and $\mathscr{D}[\mathbf{\Omega}]$ is the infinitesimal ``volume'' element in the $\mathbf{\Omega}$ space. 

As described in the main text, the probability-weight $P(\{\Omega_{i}\})$ of finding the system with a characteristic $\Omega_{i}$ and energy $E$, is obtained by integrating out Eq.~\ref{eq.app.prob} along the two other local deformations. Then, for $\Omega_{3}$ we get the following relation:

\begin{equation}
\begin{aligned}
P(\{ \Omega_3 \}) =&\frac{1}{Z} \mathlarger{\prod_{n=1}^{N} \iint} e^{-\frac{\beta a}{2} [A_1\Omega_1^2 + A_2(\Omega_2 + \frac{G}{A_2} \Omega_3)^{2} + {\tilde C} \Omega_{3}^{2}  ]} d\Omega_{1} d\Omega_{2}\\
  =&\frac{1}{Z} \left( \frac{4\pi^2}{\beta^2 a^2 A_1 A_2} \right)^{N/2} \mathlarger{\prod_{n=1}^{N}} e^{-\frac{\beta a}{2} {\tilde C} \Omega_3^2}.
\end{aligned}
\label{pomega3}
\end{equation}

\noindent The analogous calculation for $\Omega_{2}$ gives:

\begin{equation}
P(\{ \Omega_2 \}) = \frac{1}{Z} \left( \frac{4\pi^2}{\beta^2 a^2 A_1 C} \right)^{N/2} \mathlarger{\prod_{n=1}^{N}} e^{-\frac{\beta a}{2} {\tilde A_2} \Omega_2^2}.
\label{pomega2}
\end{equation}

\noindent Therefore, the rescaled constants ${\tilde C} = C \left( 1- \frac{G^2}{A_2 C}\right)$ and ${\tilde A}_2 = A_2 \left( 1- \frac{G^2}{A_2 C}\right)$ that appear in the exponential functions above, imply the softening of the twisting and bending response, respectively. 

One additional observation is that the average $\langle \mathscr{O} \rangle_{\Omega_3}$ of any observable $\mathscr{O}$ at a constant value of $\Omega_{3}$ is found through the following equation:
\begin{equation}
\langle \mathscr{O} \rangle_{\Omega_3} = \frac{1}{Z} \mathlarger{\int} \mathscr{D}[\Omega_1, \Omega_2] \mathscr{O} e^{-\beta E}.
\label{eq.app.average}
\end{equation}
Therefore, the averages $\langle \Omega_{1} \rangle_{\Omega_3}$ and $\langle \Omega_{2} + \frac{G}{A_{2}} \Omega_{3} \rangle_{\Omega_3}$ are found to be zero. This key result implies that the internal torques of the molecule, $M_1$ and $M_2$, are also zero.

Finally, it is worth noting that the calculations presented here can be generalized to the ones of a ring molecule by replacing the energy of the system (Eq.~(\ref{e_complete_square})) by the appropriate one (Eq.~(\ref{eq.energydensity})), and rewriting it in terms of the deviations $\delta \Omega_{i}(s)=\Omega_{i}(s)-\Omega_{i}^{0}(s)$ of the deformations $\Omega_{i}(s)$, in a similar way to what is done at the beginning of the next section. The outcome of this approach is that we obtain similar equations to the ones displayed here but with $\Omega_i$ replaced by  $\delta \Omega_{i}$: $\langle \delta \Omega_{1} \rangle_{\delta\Omega_3}=0$ and $\langle \delta \Omega_{2} + \frac{G}{A_{2}} \delta \Omega_{3} \rangle_{\delta\Omega_3} = 0$.

\section{\label{App.kirchoff} Internal torque components}
As described in the main text, the components of the internal torque are found by computing the derivative of the energy density Eq.~(\ref{eq.energydensity}) with respect to the elastic strain: $M_{i} = \frac{\delta \varepsilon}{\delta \Omega_{i}}$. Here we write the results in terms of the deviations $\delta \Omega_{i}(s)=\Omega_{i}(s)-\Omega_{i}^{0}(s)$ of the deformations $\Omega_{i}(s)$ with respect the minimum energy state $\Omega_{i}^{0}(s)$ (given in Eq.~(\ref{eq.groundstate})).

Using $\delta \Omega_i(s)$, the energy density is rewritten as:
\begin{eqnarray}
\varepsilon(s) &=& \frac{A_1}{2} \delta \Omega_1^2
+\frac{ A_2}{2} \delta \Omega_2^2 + \frac{C}{2} \delta \Omega_3^2 + G \delta \Omega_2 \delta \Omega_3 \nonumber \\
&& + \epsilon_G(s),
\end{eqnarray}
where $\epsilon_G(s) \equiv \varepsilon(\Omega_{1}^{0}, \Omega_{2}^{0}, \Omega_{3}^{0})$ represents the ground state energy density, thus, independent of $\delta \Omega_i$. Note that the ground energy density is
\begin{eqnarray}
\epsilon_G(s) &=& -\frac{\mu^2}{2A_1}\sin^2{(\chi s)} - \frac{\mu^2}{2{\tilde A}_2}\cos^2{(\chi s)}\nonumber \\
&&- \frac{\lambda^2}{2 {\tilde C}} + \mu \lambda \frac{G}{{\tilde A}_2 C}\cos{\chi s}, 
\end{eqnarray}
and its contour integral 
\begin{eqnarray}
\int_G^L \epsilon_G(s) \ ds = - \left(\frac{k_BT }{2}\frac{l_b}{R_0^2} + \frac{{\tilde C} }{2}\Delta \chi^2 \right)
L,
\end{eqnarray}
is formally identical to the energy of torsionally stressed ring (with radius of curvature $R_0$ and the average excess twist density $\Delta \chi$) made from isotropic TWLC with the bending and twisting moduli ${\tilde A}$ and ${\tilde C}$, respectively~\cite{2019Ecarlonpolygonalshapes}. Since $\partial \varepsilon/\partial \Omega_i = \partial \varepsilon/\partial (\delta \Omega_i)$, we find
\begin{eqnarray}
M_{1}&=&A_{1}\delta \Omega_{1}, \\
M_{2}&=&A_{2} \delta \Omega_{2} + G \delta \Omega_{3}, \\
M_{3} &=& C\delta \Omega_{3} + G \delta \Omega_{2}.
\end{eqnarray}

The tangential component of the derivative of the internal torque ($\mathbf{M}=M_{1}\uvec{e}_{1}+M_{2}\uvec{e}_{2}+M_{3}\uvec{e}_{3}$) with respect $s$ is obtained by using the relation in Eq.~(\ref{eq.deds}) and by  noticing that the only non-vanishing terms are the following:
\begin{equation}
\begin{aligned}
\frac{d \mathbf{M}}{ds} \cdot \uvec{e}_{3} =&{} \left[ M_{1} \frac{d\uvec{e}_{1}}{ds} + M_{2} \frac{d\uvec{e}_{2}}{ds} \right] \cdot \uvec{e}_{3} + \frac{d M_{3}}{ds}\\
     =&{}M_{2}\Omega_{1} - M_{1}\Omega_{2} + \frac{d M_{3}}{ds}\\
     =&{}A_{2} \Omega_{1} \delta \Omega_{2} - A_{1} \Omega_{2} \delta \Omega_{1} + G \Omega_{1} \delta \Omega_{3} +\\
      &{} \frac{d}{ds} (C \delta \Omega_{3} + G \delta \Omega_{2}).
\end{aligned}
\end{equation}

\section{\label{App.comprelations} Compatibility relation}
The derivative of the local reference frame with respect to the intrinsic length, $s$, and time, $t$, are expressed by the relation:

\begin{equation}
\frac{\partial \uvec{e}_{\alpha} }{\partial s}  = \mathbf{\Omega}_{\rm T}(s,t) \times \uvec{e}_{\alpha}(s,t),
\label{SI_B_spatial_derivative}
\end{equation}

\begin{equation}
\frac{\partial \uvec{e}_{\alpha} }{\partial t}  = \boldsymbol{\omega}(s,t) \times \uvec{e}_{\alpha}(s,t).
\label{SI_B_time_derivative}
\end{equation}

Compared to Eq.~(\ref{eq.deds}) we have simplified notation in Eq.~(\ref{SI_B_spatial_derivative}) by defining a total strain vector $\mathbf{\Omega}_{\rm T}(s) = \mathbf{\Omega}(s) + \chi_0(s) \uvec{e}_3(s)$, where $\chi_0$ is the intrinsic twist rate.
Now if we consider the combined action of space and time on the reference frame, since $s$ and $t$ are independent variables, they must commute and we could write the equation:

\begin{align}
0  =& \frac{\partial}{\partial t} \frac{\partial \uvec{e}_{\alpha} }{\partial s} - \frac{\partial }{\partial s} \frac{\partial \uvec{e}_{\alpha} }{\partial t} \nonumber \\
   =& \frac{\partial}{\partial t} [ \mathbf{\Omega}_{\rm T}\times \uvec{e}_{\alpha}] - \frac{\partial }{\partial s} [ \boldsymbol{\omega} \times \uvec{e}_{\alpha} ] \nonumber \\ 
   =& \frac{\partial \mathbf{\Omega}_{\rm T}}{\partial t}  \times \uvec{e}_{\alpha} + \mathbf{\Omega}_{\rm T} \times \frac{\partial \uvec{e}_{\alpha}}{\partial t}  - \frac{\partial \boldsymbol{\omega} }{\partial s}  \times \uvec{e}_{\alpha} - \boldsymbol{\omega} \times \frac{\partial \uvec{e}_{\alpha} }{\partial s} \nonumber \\
   =& \left[ \frac{\partial \mathbf{\Omega}_{\rm T}}{\partial t} - \frac{\partial \boldsymbol{\omega} }{\partial s} \right] \times \uvec{e}_{\alpha} + \mathbf{\Omega}_{\rm T} \times \frac{\partial \uvec{e}_{\alpha}}{\partial t}  - \boldsymbol{\omega} \times \frac{\partial \uvec{e}_{\alpha} }{\partial s}  \nonumber \\
   =& \left[ \frac{\partial \mathbf{\Omega}_{\rm T}}{\partial t} - \frac{\partial \boldsymbol{\omega} }{\partial s} \right] \times \uvec{e}_{\alpha} + \mathbf{\Omega}_{\rm T} \times [ \boldsymbol{\omega} \times \uvec{e}_{\alpha} ] \nonumber \\
    & -\boldsymbol{\omega} \times [ \mathbf{\Omega}_{\rm T} \times \uvec{e}_{\alpha} ]. \label{SI_B_substitution2}
\end{align}

Using the property of the cross product $a \times (b \times c) = b( a \cdot c ) - c (a \cdot b)$ on the last two terms we get 

\begin{align*}
\left[ \frac{\partial \boldsymbol{\omega}}{\partial s} - \frac{\partial \mathbf{\Omega}_{\rm T}}{\partial t} \right] \times \uvec{e}_{\alpha}   = &  \boldsymbol{\omega} [\mathbf{\Omega}_{\rm T} \cdot \uvec{e}_{\alpha} ]  - \uvec{e}_{\alpha} [ \mathbf{\Omega}_{\rm T} \cdot \boldsymbol{\omega} ] - \mathbf{\Omega}_{\rm T} [ \boldsymbol{\omega} \cdot \uvec{e}_{\alpha} ] \\
& + \uvec{e}_{\alpha} [ \boldsymbol{\omega} \cdot \mathbf{\Omega}_{\rm T} ]  \\
=& \boldsymbol{\omega} [\mathbf{\Omega}_{\rm T} \cdot \uvec{e}_{\alpha} ]  - \mathbf{\Omega}_{\rm T} [ \boldsymbol{\omega} \cdot \uvec{e}_{\alpha}(s,t) ] \\
=& \boldsymbol{\omega} [ \uvec{e}_{\alpha} \cdot \mathbf{\Omega}_{\rm T}]  - \mathbf{\Omega}_{\rm T} [ \uvec{e}_{\alpha} \cdot \boldsymbol{\omega}(s,t)  ] \\
=& \uvec{e}_{\alpha} \times [ \boldsymbol{\omega} \times \mathbf{\Omega}_{\rm T} ] \\
=& - [ \boldsymbol{\omega} \times \mathbf{\Omega}_{\rm T}] \times \uvec{e}_{\alpha}.
\end{align*}

As the above relation holds for any component $\alpha$, we get the compatibility relation:
\begin{eqnarray}
\frac{\partial \mathbf{\Omega}_{\rm T}}{\partial t } - \frac{\partial \boldsymbol{\omega}}{\partial s} - \boldsymbol{\omega} \times \mathbf{\Omega}_{\rm T} = 0. \label{SI_B_PreliminarCompatibilityEquation} 
\end{eqnarray}
Using the fact that the intrinsic twist rate $\chi_0$ is independent of time, one can rewrite the above equation into the following form:
\begin{eqnarray}
     \frac{\partial \mathbf{\Omega}(s,t)}{\partial t}  = \frac{\partial \boldsymbol{\omega}(s,t)}{\partial s} + \boldsymbol{\omega}(s,t) \times \mathbf{\Omega}(s,t). \label{SI_B_CompatibilityEquation1}
\end{eqnarray}

The dependence on the reference frame could be worked out 

\begin{align*}
         \frac{\partial {\Omega}_\alpha}{\partial t} \uvec{e}_{\alpha} + {\Omega}_\alpha \frac{\partial \uvec{e}_{\alpha}}{\partial t}  & = \frac{\partial {\omega}_\alpha}{\partial s} \uvec{e}_{\alpha} + {\omega}_\alpha \frac{\partial \uvec{e}_{\alpha}}{\partial s}  + \boldsymbol{\omega} \times \mathbf{\Omega}\\
         \frac{\partial {\Omega}_\alpha}{\partial t} \uvec{e}_{\alpha} + \boldsymbol{\omega} \times \mathbf{\Omega} & = \frac{\partial {\omega}_\alpha}{\partial s} \uvec{e}_{\alpha} + \mathbf{\Omega}_{\rm T} \times \boldsymbol{\omega} + \boldsymbol{\omega} \times \mathbf{\Omega}\\
         \frac{\partial {\Omega}_\alpha}{\partial t} \uvec{e}_{\alpha} & = \frac{\partial {\omega}_\alpha}{\partial s} \uvec{e}_{\alpha} + \mathbf{\Omega}_{\rm T} \times \boldsymbol{\omega}\\
\end{align*}
From this equation it is immediate to show the projection along the $\uvec{e}_{3}$ axis:
\begin{equation}
     \frac{\partial \Omega_3}{\partial t}  = \frac{\partial \omega_3}{\partial s} + \Omega_1 \omega_2 - \Omega_2 \omega_1 
     \label{SI_B_CompatibilityEquation}
\end{equation}

The angular velocity $\omega_3$ on the left could be obtained from the torque balance equation, while the last term is related to the writhe and could be rewritten as   $\Omega_1 \omega_2 - \Omega_2 \omega_1 = (\uvec{e}_3 \times \partial_s \uvec{e}_3) \cdot \partial_t \uvec{e}_3$

\section{\label{App.supercoiling}DNA supercoiling}
For closed DNA molecules the number of times that the two strands winds around each other (the linking number $\mathrm{Lk}$) is a topological invariant. Further more, under this circumstances the well known White-Fuller-Calugarenau theorem~\cite{1978Fuller} must be satisfied. This theorem states that the linking number can be expressed as the sum of two quantities: twist ($\mathrm{Tw}$) and writhe ($\mathrm{Wr}$). The former represents the extent of rotation of the two strands around the DNA axis and the latter represents the number of self-crossings of the DNA centerline.

The DNA double helix has a preferred configuration where the two strands wrap around each other approximately once every 10.5 base pairs. In this configuration the linking number has a characteristic value $\mathrm{Lk}_{0} \simeq N/10.5$. A DNA molecule whose linking number differs from the one in the relaxed state is said to be supercoiled. Therefore, the superhelical density:

\begin{equation}
\sigma = \frac{ \Delta \mathrm{Lk}}{\mathrm{Lk}_{0}} = \frac{\mathrm{Lk} - \mathrm{Lk}_{0}}{\mathrm{Lk}_{0}},
\label{eq.shd}
\end{equation}
 
\noindent is a quantitative measure of DNA supercoiling. In this manuscript we work mainly with molecules that are 312 bp long. Therefore we expect $\mathrm{Lk}_{0}=30$. While the undertwisted molecule is initialized with $\mathrm{Lk}=29$, the overtwisted molecule has $\mathrm{Lk}=31$. Therefore we expect for these cases a small level of supercoiling $\lvert \sigma \rvert = 0.033$.

\section{\label{App.elasticpar}Elastic parameters}
The elastic parameters of the oxDNA model have been thoroughly characterized in reference~\cite{2017EnricoEoGA}. There, the authors found that the local stiffness parameters associated to the deformation at the single base-pair level ($m=1$ data in supplementary Fig. S3 of ~\cite{2017EnricoEoGA} and also reported in  SM of \cite{2018Enricotwistwavesnuc}) are: for oxDNA1 $\beta A_{1}=51$ nm, $\beta A_{2}=30$ nm, $\beta C=77$ nm and $G=0$ while for oxDNA2 $\beta A_{1}=51$ nm, $\beta A_{2}=37$ nm, $\beta C=74$ nm and $\beta G=22$ nm. It should be stressed here that the $\beta$ dependence of these parameters (that does not appear explicitly in the cited references) comes from our choice of notation when defining the free energy of the system. By using these values, the rescaled persistence length ($\beta {\tilde A}_{I}=37.8$ nm, $\beta {\tilde A}_{II}=38.14$ nm), bending rigidity about $\uvec{e}_2$ ($\beta \tilde{A}_{2I}=30$ nm, $\beta \tilde{A}_{2II}=30.46$ nm) and torsional stiffness ($\beta {\tilde C}_{I}=77$ nm and $\beta {\tilde C}_{II} \simeq 61$ nm) can be found for both models: oxDNA1 (I) and oxDNA2 (II). The ratio ${\tilde C}_{II}/{\tilde C}_{I}=0.791$ is used in the main text when comparing the diffusion coefficient of both models. We also used $\tilde{C}_{I}/{\tilde A}_{I} = 2.04$ and $\tilde{C}_{II}/{\tilde A}_{II} = 1.6$ in order to compute the bending modes (Eq.~(\ref{eq.freqdeformation})).

\begin{figure*}[htpb]
\centering
\includegraphics[width=1.0\textwidth]{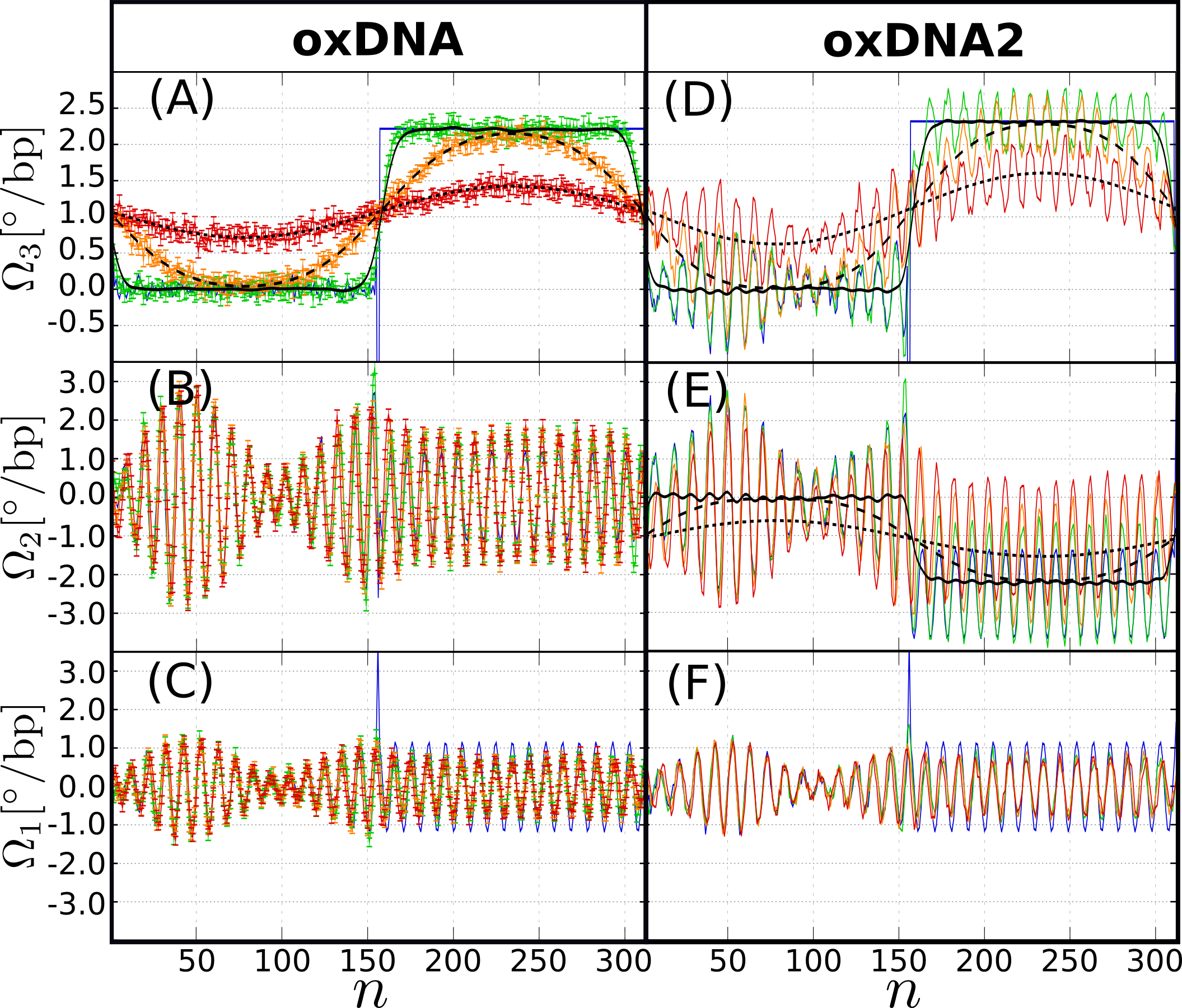}
\caption{Time evolution of the local deformations for the overtwisted oxDNA models in the simulation with planes captured at different timesteps: starting configuration (blue), 50 $\tau_{LJ}$ (green), 2500 $\tau_{LJ}$ (orange) and 15000 $\tau_{LJ}$(red). The values represent the variation of the respective deformation from the expected average (at $t=0$) in the non-overtwisted region. Left and right panels show results for the oxDNA I and II models respectively. Color dots represent data computed from simulations and color lines are a guide for the eye. To ease the visualization, error bars are only reported for left panels. However, the size of the errorbars for right panels is similar. (\textbf{A}) and (\textbf{D}) show the average local twist deformations for the two models. As consequence of the coupling ($G>0$) the right panel displays twist waves. Solid, dashed and dotted black lines represent the fit of the data to Eq.~(\ref{eq.twdiffusion}) at the corresponding timesteps. This shows that twist follows a diffusive pattern in agreement with the theory presented in this manuscript. (\textbf{B})-(\textbf{C}) are the two bending deformations $\Omega_1$, $\Omega_2$ for oxDNA1. The presence of waves with a periodicity equal to the DNA pitch is a consequence of the helical structure of DNA. Notably there are no significant changes in these variables during the twist diffusion stage, in agreement with the theory for $G=0$. (\textbf{E})-(\textbf{F}) Bending deformations for oxDNA2. The helical structure of DNA also generates waves but this time, due to the coupling between $\Omega_{2}$ and $\Omega_{3}$ the overtwisted region of $\Omega_2$ shows a clear shift consistent with Eq.~(\ref{eq.groundstate}). Furtheremore, $\Omega_2$ is also described by a diffusion equation with a very similar diffusion coefficient as $\Omega_{3}$. From the fit to the data we obtain for the overtwisted case $D_{II}^{+}(\Omega_2)=0.19\pm0.03\; bp^{2}/\tau_{\text{LJ}}$ and for the undertwisted case $D_{II}^{-}(\Omega_2)=0.20\pm0.04\; bp^{2}/\tau_{\text{LJ}}$. Solid, dashed and dotted black lines represent this fit to the diffusion equation at the same timesteps as the ones depicted in (D).}
\label{fig:paneldefevolution}
\end{figure*}

\section{\label{App.details}Details of simulations}
The Langevin integration of the system was carried out in the most recent implementation~\cite{oxDNALAMMPS} of the oxDNA model into the LAMMPS~\cite{LAMMPS} (Large Scale Molecular Massively Parallel Simulator) engine. Briefly, this model describes DNA at the single nucleotide level by means of a rigid body with additive-pairwise interaction sites. The potentials involved in the interactions accurately represent: the hydrogen bonding between complementary bases, the connectivity of the sugar-phosphate backbone, the excluded volume between nucleotides and also the stacking, coaxial-stacking and cross-stacking forces. Hence, if $U$ represents the total potential field experienced by the nucleotides and $\mathbf{r}$ their position, then the system obeys the Langevin-equation:

\begin{equation}
m\frac{d^{2} \mathbf{r}}{dt^{2}}= -\xi \frac{d \mathbf{r}}{dt} - \nabla U + \sqrt{2k_{B}T\xi} \Lambda(t),
\label{eq.langevin}
\end{equation}

\noindent where $m$ is the mass of the nucleotide, $\xi$ is the friction and $\Lambda(t)$ is the white noise term with zero mean which satisfies $\langle \Lambda_{\alpha}(t) \Lambda_{\beta}(s) \rangle = \delta_{\alpha \beta} \delta(s-t)$ along each Cartesian coordinate represented by the Greek letters. The form of the last term in Eq.~(\ref{eq.langevin}) ensures that the equipartition theorem is satisfied.\newline

\textbf{Mapping of simulation units --} The relation between one simulation unit (SU) in the oxDNA code and the international system (SI) units, is the following: mass ($M=100\text{AMU}=1.66\times10^{-25}$ kg), temperature ($T=3000$ K), length ($L_{s}=8.518\times10^{-10}$ m), energy ($\varepsilon_{s}=k_{B}T=4.142\times10^{-20}$ J) and force ($F=\varepsilon_{s}/L_{s}=4.863\times10^{-11}$ N). The simulation time $\tau_{LJ}=L_{s}\sqrt{M/\varepsilon_{s}}=1.7$ ps, comes naturally from the above quantities and it is employed to define a constant integration timestep $\Delta t = 0.001\tau_{LJ}$ of the Langevin equation (\ref{eq.langevin}). In principle, $\tau_{LJ}$ could be used to compare results with experiments. However, since the hydrodynamic effects are neglected in the Langevin formalism, one needs to be cautious in interpreting time units in this type of coarse-grained simulations.

It is also important to recognize that there are two further time scales in the system with an intuitive physical meaning. One is the inertial time $\tau_{in}=m/\xi$, which gives the characteristic time after which the velocity of a bead becomes uncorrelated. The second is the Brownian time $\tau_{Br}=(2r_{0})^{2}/D$, which gives the order of magnitude of the time it takes for a bead to diffuse across its own diameter ($2r_{0}$). Here $D$ is the translational diffusion constant for a bead, given through the Einstein relation by $D=k_{B}T/\xi$. In the approximation in which a nucleotide diffuses like a sphere with radius $r_{0}=1$ nm, we can use Stokes' law $\xi = 6\pi\eta r_{0}$, where $\eta$ is the viscosity of the fluid. Therefore, setting the values of $D$ (or similarly the value of $\eta$) and $m$, will resolve the two additional timescales. For example, if we consider that the mass of individual nucleotides is ($m=315.75\text{AMU}=5.24\times10{-25}$ kg) and they are immerse in water ($\eta=1.1\times10^{-3}\text{kg}\text{m}^{-1}\text{s}^{-1}$), we find $\tau_{in}=5.05\times10^{-14}\text{s}\simeq 0.03 \tau_{LJ}$ and $\tau_{Br}=2.5\times10^{-10}\text{s}\simeq 150 \tau_{LJ}$; with the timescales separated by several orders of magnitude ($\tau_{in} \ll \tau_{LJ} \ll \tau_{Br}$).

As pointed out in references~\cite{Tom-thesis,2013Doye}, due to the limitations in our calculations when neglecting the hydrodynamic effects, the diffusion coefficient (and then also $\tau_{in}$) could be seriously underestimated. Therefore, we need to bear in mind the timescales of interest in our system, before choosing the magnitude of $D$. For instance, in order to investigate the fast process of twist diffusion (not to be confused with $D$), which occurs at short time-scales, we use the default value of the inertial time ($\tau_{in}=0.03$) given in the original parametrization of the model. On the other hand, the supercoiling of the molecule occurs at a much larger time-scale. Setting such a low inertial time would lead to prohibitively slow writhing dynamics and unfeasibly long simulations. Instead we chose larger diffusion coefficients (see section \ref{App.totaltwist} for details) such that $\tau_{in}\leq \tau_{LJ}\leq \tau_{Br}$. This assumption means that bodies have more inertia than in reality and that processes which occur on time-scales below the Brownian time are not resolved accurately, however this is of no practical consequence for our purpose. 

It should be emphasized here that this artificial change of the diffusion, makes difficult to map the simulation time onto real units. We then prefer to report our results in units of $\tau_{LJ}$ and to focus on the comparison of times between similar processes.\newline

\textbf{Additional features of the MD simulations --} In the simulations, a ring molecule of $N=312$ bp was initialized with a deficit of twist. However, when not set properly, undertwisting encourages the local melting of the base-pairs, creating small regions where the dsDNA splits into its two single-strand components and therefore where the local twist can not be defined. To avoid this, we set appropriate physiological and geometrical conditions: (i) we used a High salt concentration of $[Na^{+}]=1M$. The Debye Huckel potential implemented in the oxDNA model allows to effectively modulate the electrostatic interaction of the nucleotides by setting the salt concentration of the system. A high value corresponds to the screening of the negatively charged phosphates of DNA, which prevents melting. (ii) We simulated poly-C molecules. Because G-C pairs form three hydrogen bonds, while A-T pairs form only two, the hydrogen bonding energy of the former is larger than the latter in the oxDNA model. Therefore we use a DNA sequence made of only G-C pairs (homopolymer). As discuss in the main text, this also ensures that the elastic parameters ($G, C, A_{1}$ and $A_{2}$) do not depend on the position ($s$) along the dsDNA. (iii) We set a low level of supercoiling in the initial configuration. Under no torsional stress the total twist of a 312 bp ring molecule is $\mathrm{Tw_{0}=30}$. At $t=0$ we start from a  conformation with $\mathrm{Tw}=29$ and $\mathrm{Wr}=0$. This corresponds to a supercoiling $\sigma=\Delta\mathrm{Tw}/\mathrm{Tw}_{0} =- 0.033$. This deficit was split among half of the ring so the local twist deficiency is small enough to avoid melting. This choice also discourages the formation of strong deviation from the planar ring configuration. (iv) Simulations were run at a low temperature ($T=15$ K). Besides favoring the hybridization of the two single strands, such a low temperature also allows to study the twist Diffusion in the absence of thermal fluctuations and in consequence less simulations have to be performed to get a good statistics. In addition, as discussed in reference \cite{2019Ecarlonpolygonalshapes}; in short, constrained and highly bent DNA, thermal fluctuations are not the main factor influencing the shape of the molecule. Finally, we also analyzed the analogous scenario for over-twisted molecules of DNA using the same conditions of the system and opposite supercoiling level ($\sigma=0.033$). As mentioned in the main text, when measuring local twist diffusion (see Fig.~\ref{fig:paneltwdiffusion}), the system was confined in between two parallel planes to avoid the writhe formation. On the other hand, when we study the evolution of total twist (see Fig.~\ref{fig:paneltwvstunder}), the planes were removed.

\section{\label{App.twistcomp}Computation of local Twist}
\begin{figure*}[htpb]
\centering
\includegraphics[width=0.92\textwidth]{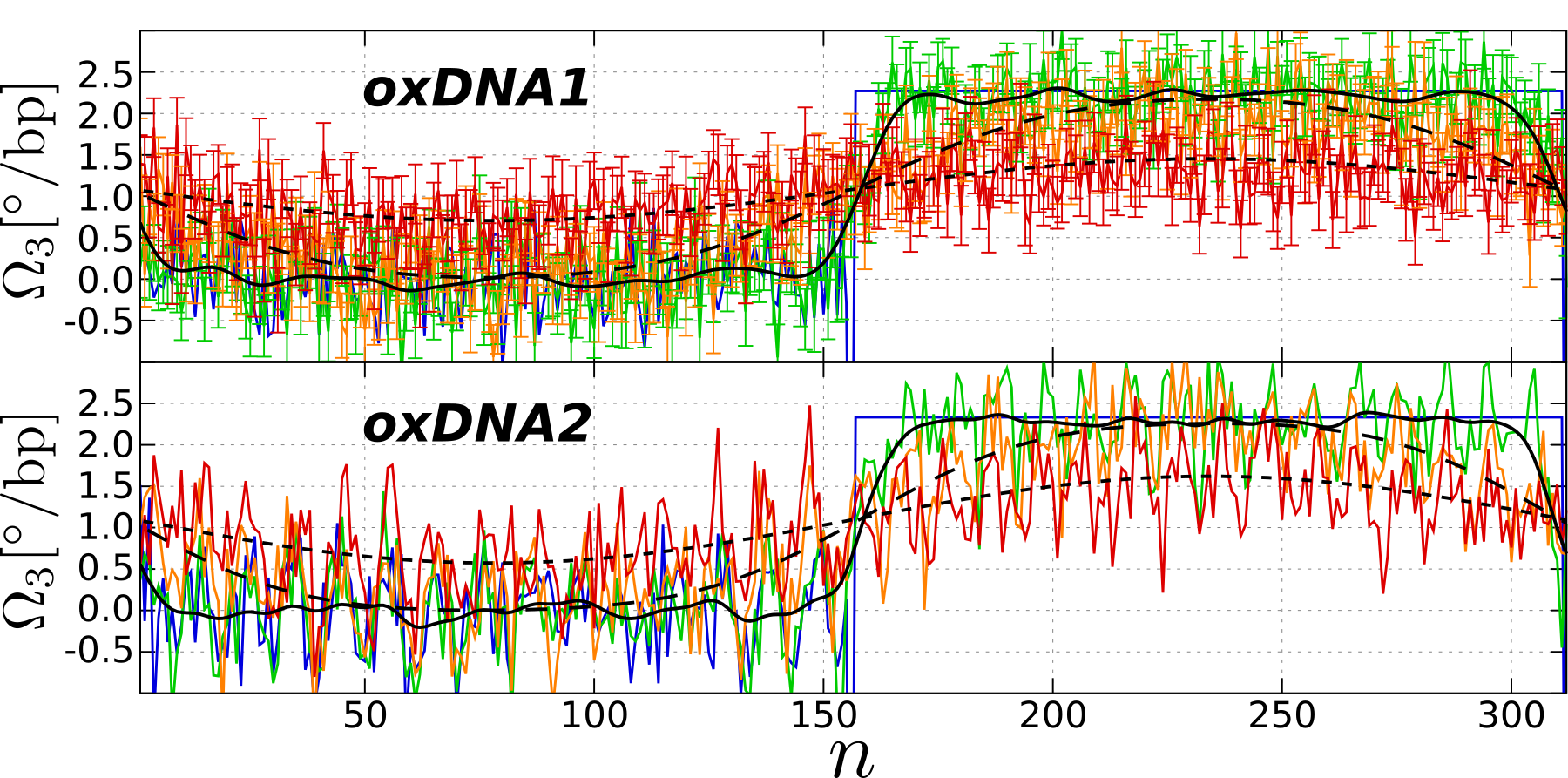}
\caption{Local twist deformation ($\Omega_{3}$) as function of the contour length when the temperature of the system is kept constant at $T=300$ K. Color lines (blue, green, yellow and red) represent results at different time-steps ($t=0, 5, 2500$ and $15000 \tau_{\text{LJ}}$).}
\label{fig:panelomega3T300K}
\end{figure*}

In order to obtain the local deformations ($\Omega_{i}$) from our simulations, we consider DNA as a discrete inextensible elastic rod. As described in reference \cite{2017EnricoEoGA}, this discretization allows to define a local reference frame $\{ \uvec{e}_{1}(n), \uvec{e}_{2}(n), \uvec{e}_{3}(n) \}$ at base-pair $n$ along the rod, using the following method. In the oxDNA model the orientation of individual nucleotides is given by two vectors: the normal to the plane of the base ($\uvec{n}$), which follows the $5^{\prime}-3^{\prime}$ direction of single strands; and the unitary vector ($\uvec{b}$), pointing from the backbone site to the base site. The intrinsic nucleotide triad is completed with a third axis defined by $\uvec{n} \times \uvec{b}$. When the DNA molecule is fully hybridized, both, the vectors $\uvec{n}_{1}$, $\uvec{n}_{2}$ that are part of the triads from two complementary nucleotides in a base-pair and their respective center of mass ($\uvec{r}_{1}$, $\uvec{r}_{2}$), become a natural choice to define the base-pair local reference frame. The tangent to the centerline of the dsDNA is $\uvec{e}_{3}=(\uvec{n}_{1}-\uvec{n}_{2})/\lvert \uvec{n}_{1}-\uvec{n}_{2}\rvert$. The vector $\uvec{e}_{2}=(\uvec{p}-(\uvec{p}\cdot\uvec{e}_{3})\uvec{e}_{3})/\lvert \uvec{p}-(\uvec{p}\cdot\uvec{e}_{3})\uvec{e}_{3} \rvert$ is defined as the projection of the vector $\uvec{p}=\uvec{r}_{1}-\uvec{r}_{2}$ connecting the two centers of mass, onto the plane perpendicular to $\uvec{e}_{3}$. The last vector is defined as $\uvec{e}_{1}=\uvec{e}_{2} \times \uvec{e}_{3}$, and it points in the direction of the symmetry axis of the DNA grooves.

The local deformations can be computed from the rotation matrix, $\mathbf{R}(n)=\mathbf{T}^{T}(n)\mathbf{T}(n+1)$, which generates the frame at segment $n + 1$ from that at segment $n$. Here the $3\times 3$ orthogonal matrix ($\mathbf{T}$), is constructed by using as columns the local reference frame vectors: $\mathbf{T}(n)=[\uvec{e}_{1}(n),\uvec{e}_{2}(n),\uvec{e}_{3}(n)]$ and its transpose is represented by $\mathbf{T}^{T}(n)$. At each position $n$ along the rod, the components $R_{ij}(n)$ of the matrix $\mathbf{R}(n)$ are related to a rotation vector $\bm{\theta}(n)=\theta_{1}(n)\uvec{e}_{1}(n) + \theta_{2}(n)\uvec{e}_{2}(n) + \theta_{3}(n)\uvec{e}_{3}(n)$ according to the following equation:

\begin{equation}
\begin{pmatrix}
\theta_{1}\\
\theta_{2}\\
\theta_{3}
\end{pmatrix}
= \frac{\theta}{2\sin\theta}
\begin{pmatrix}
R_{32} - R_{23}\\
R_{13} - R_{31}\\
R_{21} - R_{12}
\end{pmatrix},
\label{eq.rotvector}
\end{equation}

\noindent where $\theta$ satisfies the relation: $\text{trace}(\mathbf{R})=1+2\cos\theta$. Finally, the local deformations can be defined as the deviations of the components of $\bm{\theta}(n)$ from their respective mean value ($\overline{\theta_{i}}$) in the relaxed configuration (under no mechanical stress):

\begin{equation}
a\Omega_{i}(n) = \theta_{i}(n) - \overline{\theta_{i}},
\label{eq.localdef}
\end{equation}  

\noindent where $a=0.34$ nm is the mean distance between consecutive base-pairs. The values of $\overline{\theta_{i}}$ have been obtained from simulations of linear molecules in reference~\cite{2017EnricoEoGA}. For oxDNA1 it was found that $\overline{\theta_{1}} = \overline{\theta_{2}}=0$ and $\overline{\theta_{3}}=34.8\degree$, while the same quantities for oxDNA2 are: $\overline{\theta_{1}}=0$, $\overline{\theta_{2}}=2.6\degree$ and $\overline{\theta_{3}}=34.1\degree$. Note that in the main text we use $\chi_{0} = \overline{\theta_{3}}/a$ as the value of $\Omega_{3}$ in the relaxed sate of the DNA.

\textbf{Average of the local deformations --} In order to obtain the time evolution of the local deformations, we ran 100 independent configurations of the system described in section \ref{App.details} while we keep the temperature constant at $T=15$ K. All the samples start with an excess/deficit of one helical turn in half of the ring and the local twist in this region is locked during equilibration. After this stage, we release the constraint on the over/under twisted region and we keep track of the local deformations over the entire ring for $1.5\times 10^{4}\tau_{LJ}$. In practice this is done by computing the average $\langle \Omega_{i} (n) \rangle$ over configurations of the local deformations at the same time after the twist release. 

In Fig.~\ref{fig:paneltwdiffusion} of the main text we report the values obtained for $\Omega_{3} (n)$ in the undertwisted case at different times. To complement this, in Fig.~\sref{fig:paneldefevolution} we show results of the three local deformations for the overtwisted case. We stress here that during the diffusion of twist, the results for the over and undertwisted cases are essentially the same. Our results are also comparable to those on \cite{2019Ecarlonpolygonalshapes}. The simulations with the oxDNA2 model show that the twist-bend coupling generates the twist waves and the antiphase relation between $\Omega_{2}$ and $\Omega_{3}$ predicted in Eq.~(\ref{eq.groundstate}). In addition, at the beginning of the simulation, in the over twisted region the values of $\Omega_{2}$ and $\Omega_{3}$ are shifted with respect to zero by the factors $-\lambda G/ \tilde{C}A_{2}$ and $\lambda/\tilde{C}$, as expected from Eq.~(\ref{eq.groundstate}). Remarkably, the value of $\Omega_{3}(n,t)$ computed from the simulations ca be fitted perfectly by the diffusion equation (\ref{eq.twdiffusion}). Furthermore, due to the coupling, $\Omega_{2}(n,t)$ also exhibits a diffusive behavior with basically the same diffusion coefficient as the twist diffusion, in agreement with the theory presented in the main text of this manuscript.

The local deformation $\Omega_{3} (n)$, obtained for the same system as the one described above, but when the temperature is fixed at $T=300$ K, is reported in Fig.~\sref{fig:panelomega3T300K}. It is clear that by increasing the temperature of the system, the signal becomes noisier compared to the one depicted in Fig.~\sref{fig:paneldefevolution}. As pointed out in the main text, running simulations at room temperature would require to have more statistics and therefore it would be less efficient.

\section{\label{App.diffcomp}Computation of the Diffusion coefficient}
The diffusion equation (\ref{eq.twdiffusion}) has an explicit solution that can be conveniently written in terms of the Fourier coefficients. Considering the periodic boundary conditions of our system (for ring DNA) this leads to:
\begin{equation}
\delta \Omega_{3}(t,n)= \sum_{k} \tilde{W}_k(t) e^{-\frac{2\pi k}{N} n} = \sum_{k} \tilde{W}_k e^{ -\frac{4\pi^2 k^2 \tilde{D}}{N^2} t } e^{-\frac{2\pi k}{N} n} \label{eq.twdiffusion_solution}
\end{equation}
where $k \in \mathbb{Z} $ and the Fourier coefficients $\tilde{W}_k$ are obtained from the initial condition $\tilde{W}_k = \sum_{n=1}^N \delta \Omega_{3}(0,n) e^{ \frac{2\pi k}{N} n}$.
Notice that this relation holds true also in the continuum limit where $\frac{n}{N}\rightarrow s\in [0,1]$. 
If we focus our attention to a single modes $k$, we observe that comparing the Fourier coefficients at different timesteps it is possible to obtain the time-dependence:
\begin{equation}
\ln \frac{|\tilde{W}_k(0)|}{|\tilde{W}_k(t)|} = \tilde{D} \frac{4\pi^2 k^2}{N^2} t  \label{eq.diffusion_coefficient}
\end{equation}

\begin{figure}[hpt]
\centering
\includegraphics[width=\columnwidth]{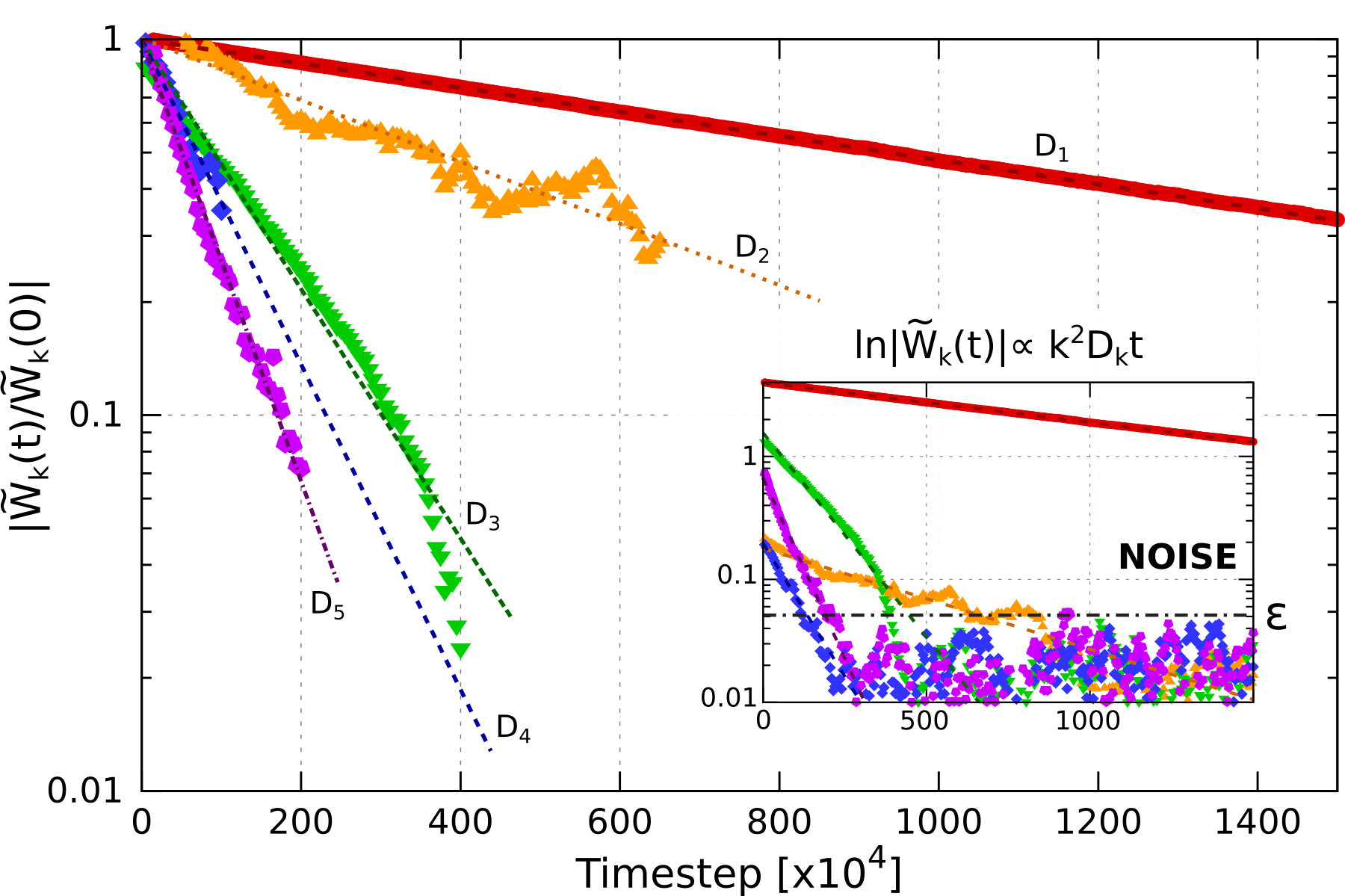}
\caption{\textbf{Calculation of the Diffusion coefficient from the Fourier Analysis}. Equation \ref{eq.diffusion_coefficient} predicts for the Fourier modes of the twist local deformation an exponential decay with time. In the main plot, time evolution of the first modes, $k=1$ red, $k=2$ yellow, $k=3$ green, $k=4$ blue, $k=5$ purple, are shown with respect to their initial values and the dashed lines represent the exponential fit. Since the noise set a limit below which no information could be extracted, the exponential are fitted in the range $t\in[0,T_\epsilon]$ where $T_\epsilon$ is the largest values for which $|\tilde{W}|_k(t)>\epsilon \; \forall \; t<T_\epsilon$. Data from all the coefficients and the level of noise $\epsilon$ are shown in the inset. Using the fit parameters it is possible to estimate the values of the diffusion constant $D_k=a_k \frac{N^2}{4\pi^2 k^2}$.}
\label{fig:DiffusionCoefficientFit}
\end{figure}

We have computed the Fourier analysis of twist for each timestep of the trajectory. Then, by using a linear fit of the Fourier coefficients via Eq.~\ref{eq.diffusion_coefficient} (see Fig.~\sref{fig:DiffusionCoefficientFit}), we were able to estimate the diffusion coefficient in the different models. The value we presented is a weighted average on the first 5 coefficients (the $k=0$ mode is obviously excluded also). In our simulation, the diffusion coefficients computed on higher modes have to be excluded because the white noise combined with faster relaxation time ($\tau \propto 1/k^2$) reduce the number of data available for the exponential fit (see Fig.~\sref{fig:DiffusionCoefficientFit} inset). The error in the diffusion coefficient is obtained from the standard error-propagation formula applied to the errors resulting from the previous fit.

In the oxDNA2 model ($G > 0$) there is a further observation to do: the twist $\Omega_3$ shows waves due to the coupling with $\Omega_2$ (see Fig.~\sref{fig:paneldefevolution}(\textbf{D})). These waves have a characteristic frequency equal to the pitch of DNA, one turn every $10.5$ base pairs. If we reconstruct this signal using only the low frequencies of the Fourier Transform of $\Omega_{3}$, we get a curve that passes through the centerline of the wave. From the fit of the diffusion equation to the data we obtain a curve that follows basically the same trajectory. The same behavior is found for $\Omega_{2}$ in agreement with our theory (see also Fig.~\sref{fig:paneldefevolution}). 

\section{\label{App.totaltwist}Evolution of the total Twist}
At any fixed timestep from the simulation, the total twist ($\mathrm{Tw}$) is found from adding the value of the local twist $\theta_{3}$ along all the base-pairs:

\begin{equation}
\mathrm{Tw} = \frac{1}{2\pi}\sum_{n=1}^{N} \theta_{3}(n).
\label{eq.totaltwist}
\end{equation}

\noindent Therefore, the sum of the twist deformations is related to the deviations of the total twist from its value under no torsional stress ($\mathrm{Tw}_{0}$) according to:

\begin{equation}
\mathrm{Tw} - \mathrm{Tw}_{0} = \frac{1}{2\pi}\sum_{n=1}^{N} a\Omega_{3}(n).
\label{eq.sumomega3}
\end{equation}

\noindent When the ring molecule is constraint by the two parallel planes, the initial writhe is preserved during the whole simulation ($\mathrm{Wr}(t)=0$) and Eq.~(\ref{eq.sumomega3}) gives a constant value of, for example, $-1$ for molecules initialized in the undertwisted case. On the other hand, when we remove the planes from the simulations the value of $\mathrm{Wr}$ changes with time. There is an exchange of twist and writhe that obeys the White-Fuller-Calugarenau theorem: $\mathrm{Lk} = \mathrm{Tw}(t) + \mathrm{Wr}(t)$, where $\mathrm{Lk}$ is constant.

\begin{figure}[htp]
\centering
\includegraphics[width=0.49\textwidth]{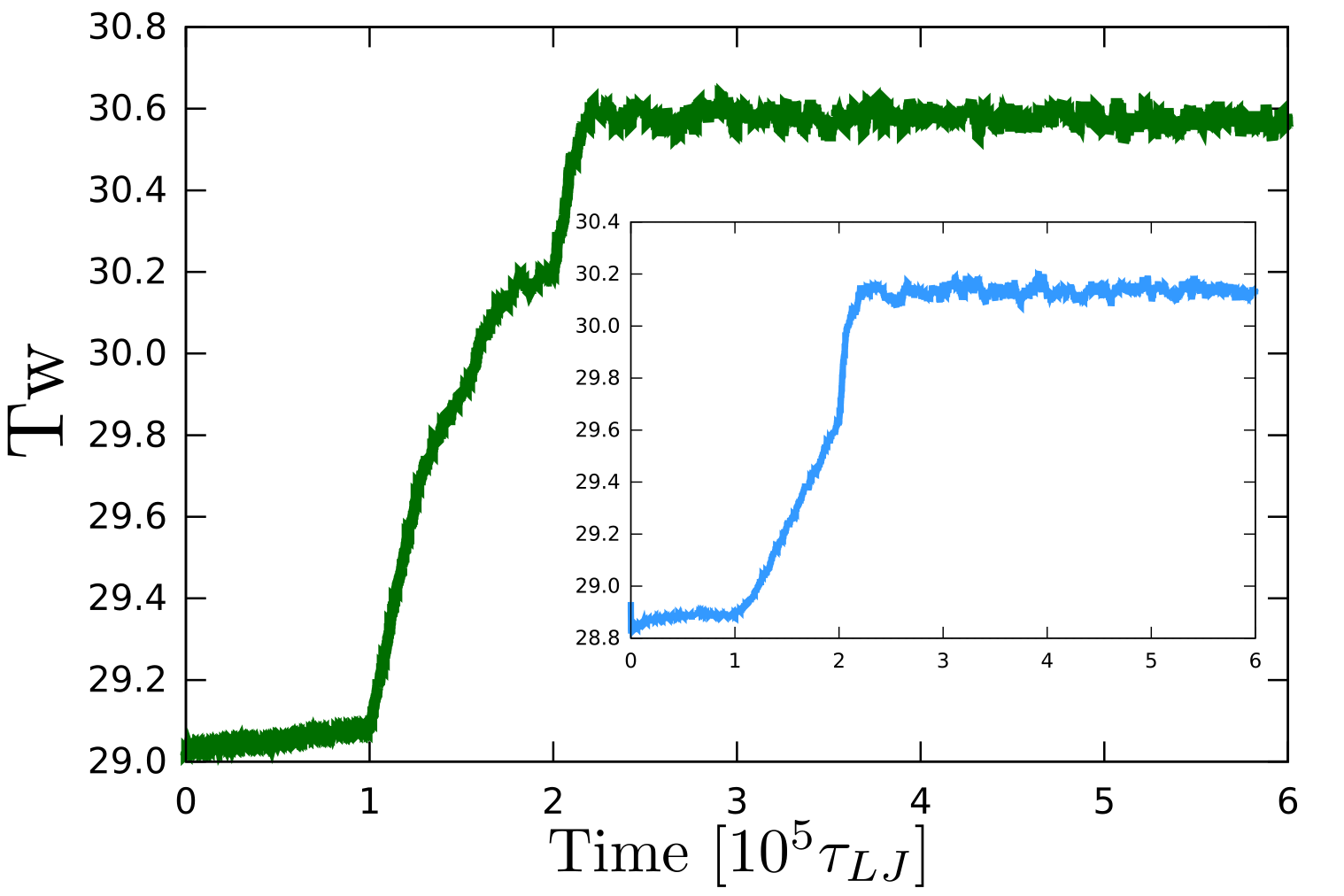}
\caption{Time evolution of the total twist for oxDNA1 (green) and oxDNA2 (cyan). Results are obtained from simulations at $T=15$ K.}
\label{fig:paneltwevolapp}
\end{figure}

The writhing of the molecule requires global conformational changes and hence is a slow process compared to the diffusion of the local twist. Therefore, to be able to track the writhe evolution (or analogously $\mathrm{Tw}(t)$) we had to speed up our simulations. The standard way of doing this in molecular dynamics simulations (see section \ref{App.details}), is to use a high diffusion constant, which in turn means to use a high inertial time $\tau_{in}$. 

We ran simulations for a total time of ($6\times10^{5}\tau_{\text{LJ}}$). Initially, the inertial time was set to $\tau_{in}=0.03\tau_{\text{LJ}}$. After $1\times10^{5}\tau_{\text{LJ}}$ we increased its value to $\tau_{in}=1\tau_{\text{LJ}}$. A second increased to $\tau_{in}=10\tau_{\text{LJ}}$ was applied at time-step $2\times10^{5}\tau_{\text{LJ}}$. The results of the total twist following this protocol and average over three independent configurations are shown in Fig.~\sref{fig:paneltwevolapp} for undertwisted ring DNA molecules when the temperature is set to 15 K. In Fig.~\ref{fig:paneltwvstunder} of the main text we report the same results after rescaling the units of time by a factor of $1/0.03 = 33.33$ during the first increment in $\tau_{in}$ and by a factor ten times larger during the second increment.

\section{\label{App.bendingmodes} Bending modes}
As explained in the main text, the dynamics of twist can be divided in two stages. At times $t<t^{*}$ the deficit of twist diffuses across the entire ring, keeping the total twist constant in the process. In other words, there is no production of writhe. At $t>t^{*}$ the buckling of the molecule might begin. The analysis for the isotropic TWLC model indicates that there is a critical value of the linking number excess  $\Delta \mathrm{Lk}  = \mathrm{Lk} - \mathrm{Lk}_{0}$, beyond which the planar configuration becomes unstable and the ring buckles and folds on itself~\cite{1985Tanaka}. According to the linear stability analysis, a characteristic frequency $\phi_m$ for the initial out of plane deformations with mode number $m$ is determined by the following equation

\begin{equation}
\phi_{m}^{2} = \frac{A}{\rho_{0}R_{0}^{4}} \displaystyle \left( f_{m} -\frac{2\pi C \Delta \mathrm{Lk}}{A}  g_{m} \right),
\label{eq.freqdeformation}
\end{equation}

\noindent where $\rho_{0}=3.3\times10^{-15} kg/m$, $R_{0}$ is the radius of the ring with constant curvature, $f_{m}=m^{4}+3m^{2}+1/2$ and $g_{m} \approx m^{3}/2$. The most unstable mode ($m^{*}$) corresponds to the minimum of the $\phi_m^2$ and it can be found as the solution to:

\begin{equation}
4m^{3} - \frac{3\pi C \Delta \mathrm{Lk}}{A}m^{2} + 6m = 0.
\label{eq.mstar}
\end{equation}
\noindent Therefore, the theory does not predict any dependence of $m^{*}$ on the total length of the molecule but only on the ratio $\frac{C}{A}$ and $\Delta \mathrm{Lk}$. We expect that this mode will grow faster than the others, and thus will be the first observed at the onset of the buckling.

\begin{figure}[t]
\centering
\includegraphics[width=0.45\textwidth]{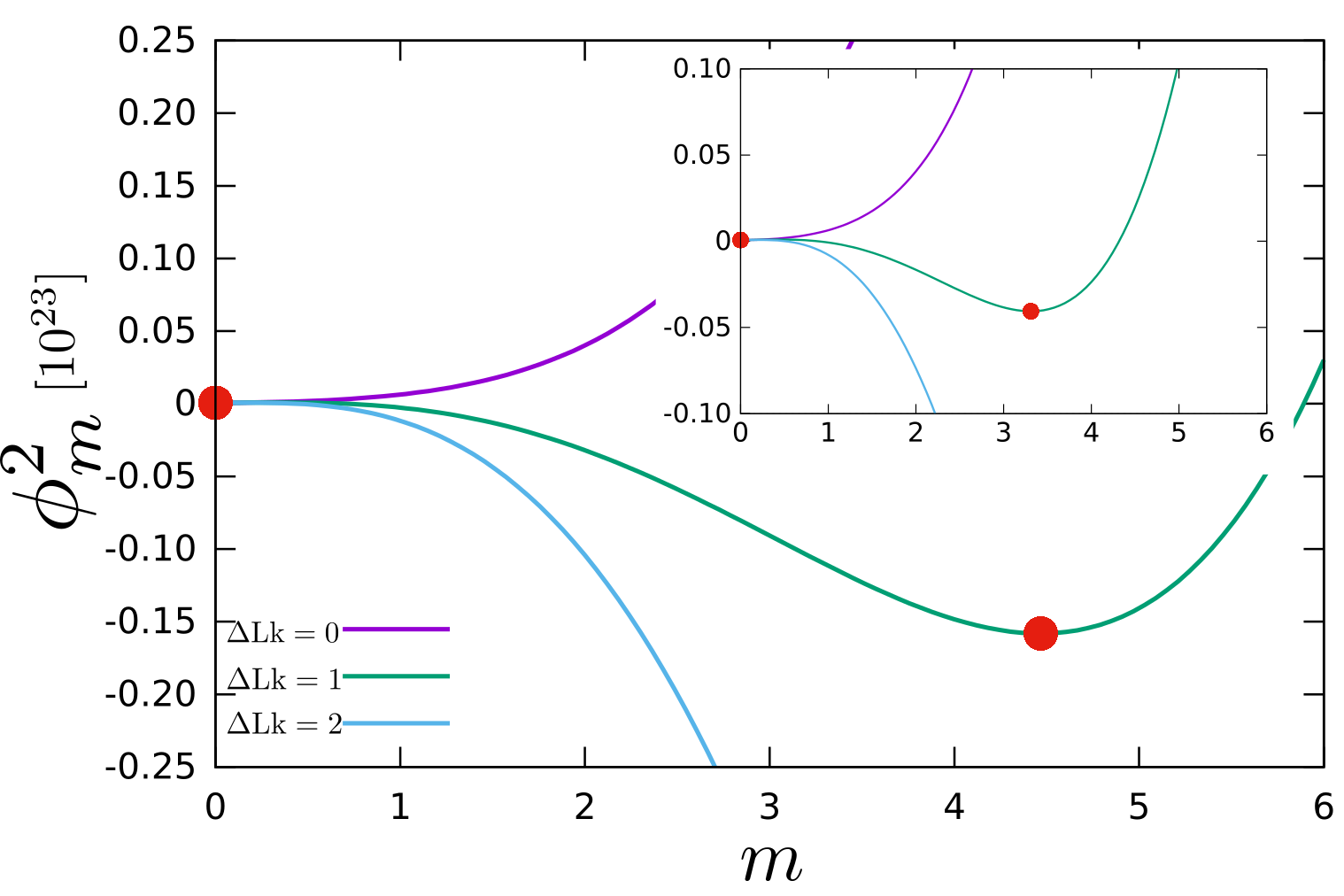}
\caption{Reduced frequencies from Eq.~(\ref{eq.freqdeformation}) against the mode number $m$, for different levels of supercoiling. At $\Delta \mathrm{Lk}=1$ the value of $\phi_{m}^{2}$ becomes negative, indicating the instability of the circle. The main figure shows results for the oxDNA1 model and the inset for oxDNA2. Minimum of each curve is depicted with a red dot.}
\label{fig:panelbendingmodes}
\vspace{-0.3cm}
\end{figure}

We now compare the prediction of the isotropic TWLC model (Eq.~(\ref{eq.freqdeformation})) with our numerical observations. By doing so, recall that the number of elastic parameters for our numerical models is larger than two. We therefore attempt to map the elastic behaviors of oxDNAs to that of an isotropic TWLC using their rescaled elastic constant ${\tilde A}$ and ${\tilde C}$. Results are displayed in Fig.~\sref{fig:panelbendingmodes}, where we plot $\phi_m^2$ for $\Delta \mathrm{Lk} = 0, 1, 2$ (analogous results are expected for negative values of $\Delta \mathrm{Lk}$, see~\cite{1985Tanaka}) using the ratio ${\tilde C}/{\tilde A}$ of oxDNA1 (2.04) and oxDNA2 (1.6). From our line of reasoning, one expects that the smaller the ratio ${\tilde C}/{\tilde A}$, the smaller the selected mode number $m^*$ at a fixed $\Delta \mathrm{Lk}$. We find that, while for oxDNA1 the minimum of $\phi_m^2$ is located at $m^* \simeq 4$, for oxDNA2 it is smaller $m^* \simeq 3$. 

\noindent Note that according to the linear stability analysis (Eq.~(\ref{eq.freqdeformation})) the value of $\Delta \mathrm{Lk}$ at which $\phi_{m}^{2}$ becomes negative (indicating the ring instability) depends on the elastic constants of the system: the smaller the ($C/A$) ratio the larger the critical $\lvert \Delta \mathrm{Lk} \rvert$ required to initiate the buckling transition. For the oxDNA models this transition happens at $\lvert \Delta \mathrm{Lk} \rvert \geq 1$. 

\begin{figure}[ht] 
\centering
\includegraphics[width=0.5\textwidth]{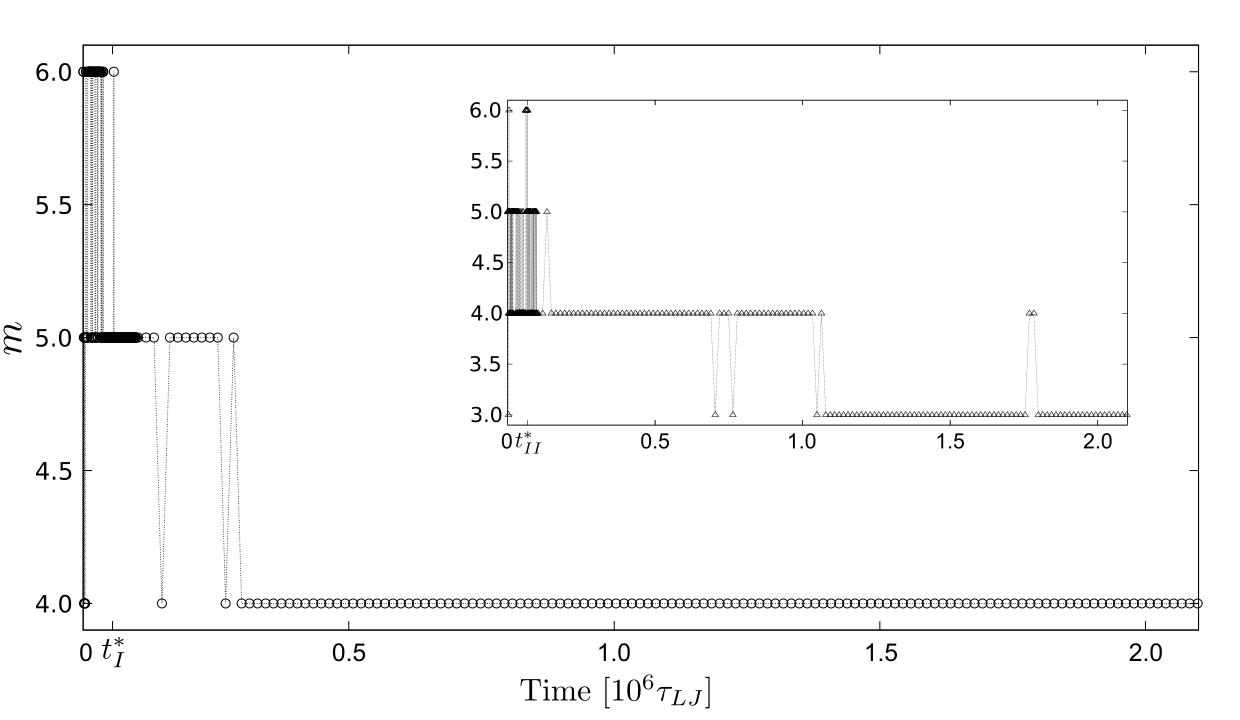}
\caption{Time evolution of the bending mode number for a molecule 312 bp long simulated with the oxDNA1 and oxDNA2 (inset) models. Lines are a guide for the eye.}
\label{app:mvst}
\end{figure}

Since the writhing of the molecule is reflected in $\Omega_{1}$ and $\Omega_{2}$ for both the oxDNA models,  we can track the bending modes ($m$) by looking at, for example, the number of minima (or maxima) that the envelope of $\Omega_{2}$ has at a certain time. This number is shown at the bottom of supplementary movies \href{run:./video/ox2_N312_deltalk1.mp4}{S1}-\href{run:./video/ox1_N312_deltalk1.mp4}{S2} and depicted as a function of time in Fig.~\sref{app:mvst}. We observe that several modes start emerging on time until one particular mode is selected: $m^{*}=4$ for oxDNA1 and $m^{*}=3$ for oxDNA2 as predicted by Eq.~(\ref{eq.mstar}) (see also Fig.~\sref{fig:panelbendingmodes}). However, it is worth mentioning here that this behavior is only true at early times. We expect that if we wait long enough until equilibration, the molecule will show the usual eight-shape (for a ring initialized with $\lvert \Delta \mathrm{Lk} \rvert = 1$) and therefore the number of modes at long times will be in general smaller than $m^{*}$. 

Since equations (\ref{eq.freqdeformation}) and (\ref{eq.mstar}) were obtained for an isotropic TWLC without twist-bend coupling, there are some features in our simulations that the theory is not able to capture. We found for example that for the oxDNA models there is always an initial increase of the selected bending mode with the ring size. This is shown in Fig.~\ref{app:mvsl} of the main text for rings with $L=312, 624$ and 936 bps and two values of linking deficit $\Delta \mathrm{Lk} =-1,-2$.

As discussed in the main text, when $G=0$, the amplitude of the oscillations in the bending deformations ($\Omega_1$ and $\Omega_2$) decreases with the size of the ring. This implies that the anisotropic case should tend to the isotropic case as $L$ becomes much larger than the persistence length. Therefore, at $L\gg l_{b}$ we should recover the no-dependence of the selected mode $m^{*}$ on the ring size. We believe that is the reason why the results for oxDNA1 with $\Delta \mathrm{Lk} =-1$ show a plateau in Fig.~\ref{app:mvsl}. We also expect that when $L\sim l_{b}$ in the anisotropic case, the larger the linking deficit the less modes observed for undertwisted rings. The net effect would be then the slow down of the growth of $m^{*}$ with $L$. Therefore, we expect that the plateau of $m^{*}$ would be reached at larger lengths as we increase the linking deficit. This is consistent with the results for oxDNA1 with $\Delta \mathrm{Lk} =-2$ in Fig.~\ref{app:mvsl} of the main text. The exact dependence of $m^{*}$ on $L$ and $\Delta \mathrm{Lk}$ is beyond the scope of this manuscript.

\section{\label{App.movies} Movies}
\newpage

\begin{figure}[h]
\renewcommand\figurename{Movie S}
\setcounter{figure}{0}    
\centering
\includegraphics[width=0.475\textwidth]{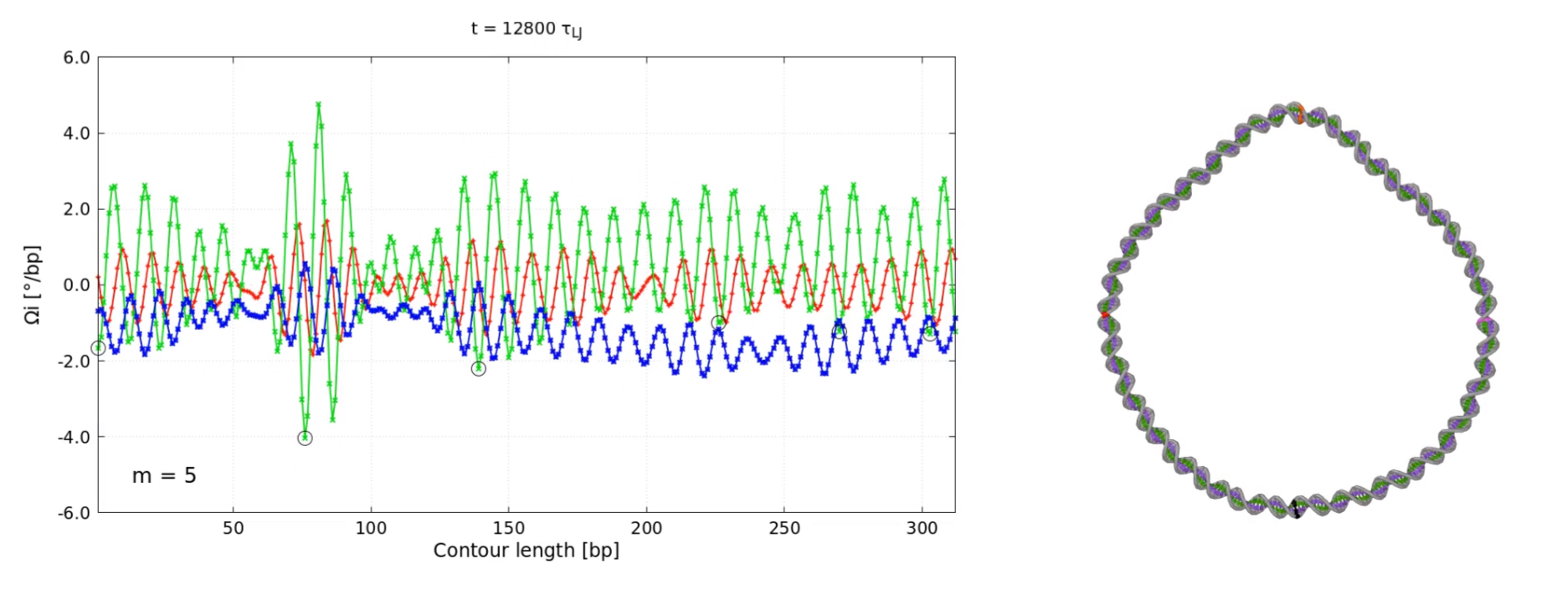}
\caption{Time evolution of the local deformations for the system simulated with the oxDNA2 model after removing the planes and releasing the twist. \textbf{Left} panel shows $\Omega_{1}$ (red), $\Omega_{2}$ (green) and $\Omega_{3}$ (blue) at the time indicated at the top of the image. The twist-bend coupling induces the twist waves. $\Omega_{2}$ and $\Omega_{3}$ are in antiphase as described in Eq.~\ref{eq.groundstate}. The local minima of the $\Omega_{2}$ envelope are depicted by black circles. Therefore, the number of black circles (considering the periodicity of the system) at a fixed time, is the number ($m$) of bending modes of the system. This is shown at the bottom of the image. \textbf{Right} panel shows the configuration of the system corresponding to the left plot. Four base-pairs located at position $n=1, 78, 156$ and 234 of the contour length are indicated with colors: pink, orange, red and black.}
\label{movies1:ox2}
\end{figure}

\begin{figure}[h]
\renewcommand\figurename{Movie S}
\setcounter{figure}{1}    
\centering
\includegraphics[width=0.475\textwidth]{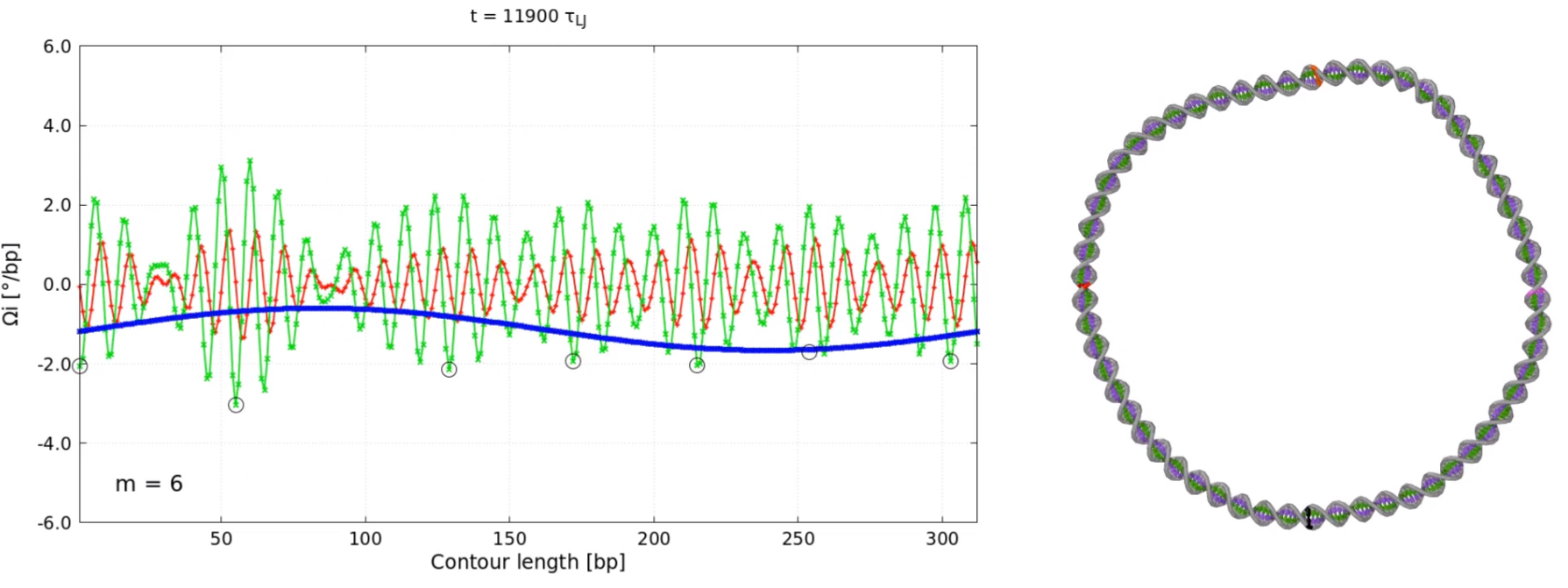}
\caption{Time evolution of the local deformations for the system simulated with the oxDNA1 model after removing the planes and releasing the twist. \textbf{Left} panel shows $\Omega_{1}$ (red), $\Omega_{2}$ (green) and $\Omega_{3}$ (blue) at the time indicated at the top of the image. The local minima of the $\Omega_{2}$ envelope are depicted by black circles. Therefore, the number of black circles (considering the periodicity of the system) at a fixed time, is the number ($m$) of bending modes of the system. This is shown at the bottom of the image. \textbf{Right} panel shows the configuration of the system corresponding to the left plot. Four base-pairs located at position $n=1, 78, 156$ and 234 of the contour length are indicated with colors: pink, orange, red and black.}
\label{movies2:ox1}
\end{figure}

\end{document}